\documentclass[aps, prl, 10pt, twocolumn, nofootinbib]{revtex4-2}
\pdfoutput=1
\usepackage{amsmath,amssymb,graphicx,xspace,subfigure,tikz}
\usepackage[dvipsnames]{xcolor}
\usepackage{bm}
\usepackage{hyperref}
\definecolor{green2}{cmyk}{0.27, 0, 1, 0.52} 
\hypersetup{
    colorlinks=true,       
    linkcolor=BrickRed,          
    citecolor=green2,        
    filecolor=magenta,      
    urlcolor=green2
}
\usepackage[all]{hypcap}
\usepackage[normalem]{ulem}
\usepackage{mathrsfs}

\newcommand{\dd}{\mathrm{d}}

\newcommand{\ii}{\mathrm{i}}
\newcommand{\xvec}{\vec{x}}

\newcommand{\nvec}{\vec{n}}

\newcommand{\LCDM}[1]{\text{$\Lambda\mathrm{CDM}$}}
\newcommand{\LCM}[1]{\text{LCM}}
\newcommand{\AMR}[1]{\text{AMR}}
\newcommand{\CMB}[1]{\text{CMB}}
\newcommand{\VOS}[1]{\text{VOS}}
\newcommand{\SM}[1]{\text{SM}}
\newcommand{\GUT}[1]{\text{GUT}}
\newcommand{\PQsc}[1]{\text{\sc pq}}
\newcommand{\PQ}[1]{\text{PQ}}
\newcommand{\UV}[1]{\text{UV}}
\newcommand{\EU}[1]{\text{EU}}
\newcommand{\gagg}{g_{a\gamma\gamma}}
\newcommand{\Healpix}[1]
{\texttt{HEALPix}}

\newcommand{\Fref}[1]{Fig.~\ref{#1}}
\newcommand{\Tref}[1]{Table~\ref{#1}}
\newcommand{\Sref}[1]{Sec.~\ref{#1}}
\newcommand{\Eref}[1]{Eq.~(\ref{#1})}
\newcommand{\Erefs}[2]{Eqs.~(\ref{#1})~and~(\ref{#2})}
\newcommand{\Rref}[1]{Ref.~\cite{#1}}
\newcommand{\Rrefs}[1]{Refs.~\cite{#1}}

\newcommand{\ie}[1]{\text{i.e.}}
\newcommand{\eg}[1]{\text{e.g.}}
\newcommand{\cf}[1]{\text{cf.}}

\makeatletter
\newcommand{\brickemail}[1]{%
  \@AF@join{\href{mailto:#1}{\textcolor{BrickRed}{#1}}}}
\makeatother

\begin{document}

\title{
CMB Birefringence from Axion String Networks Calibrated to an AMR Simulation
}

\author{Mustafa A.~Amin}

\author{Mudit Jain}

\author{Andrew J.~Long}

\author{Aden Pugsley}

\author{Moira Venegas}
\brickemail{mv58@rice.edu}

\author{Magdalena Whelley}

\affiliation{Department of Physics and Astronomy, Rice University, Houston, Texas 77005, U.S.A.}

\begin{abstract}
A cosmological network of axion strings may exist in the Universe today. If axion-like particles couple to electromagnetism, such a network induces spatially varying birefringence in the polarization of the cosmic microwave background (CMB), which can be probed by current and next-generation CMB experiments. We calibrate a loop-crossing model against a large-scale adaptive-mesh-refinement (AMR) simulation of axion-string network dynamics in the early Universe and use the calibrated model to predict CMB birefringence from recombination to today. We find that the non-detection of anisotropic birefringence in CMB observations places a strong upper bound on the electromagnetic anomaly coefficient $\mathcal{A}$ that enters the axion-photon coupling $g_{a\gamma\gamma} = - \mathcal{A} \alpha_\mathrm{em} / \pi f_a$.  A joint analysis of available anisotropic birefringence measurements constrains $|\mathcal{A}| < 0.24$ at 95\% C.L., which is independent of the Peccei-Quinn scale $f_a$, assuming that the axions are hyperlight so that the network survives until today. This limit strongly restricts the high-energy embedding of hyperlight axions, excluding the minimal Grand Unified Theory prediction for the electromagnetic anomaly coefficient at high significance. In addition, we discuss implications for an axion string origin of the recently reported evidence for isotropic birefringence. 
\end{abstract}

\maketitle

\paragraph*{\bf Introduction~~---\!}
Axions are predicted by many theoretically compelling extensions of the Standard Model of Elementary Particles \cite{DiLuzio:2020wdo} like String Theory \cite{Svrcek:2006yi,Arvanitaki:2010sy,Panda:2010uq,Mehta:2021pwf,Alvey:2021hjp,March-Russell:2021zfq,Gendler:2023kjt,Benabou:2023npn,Cline:2024vbd,Sheridan:2024vtt}.  
These hypothetical particles frequently arise as the spin-0 pseudo-Nambu-Goldstone boson of a spontaneously broken global symmetry \cite{Choi:2020rgn}.  
In such theories, axions are expected to interact with photons, which provides a strategy for seeking out these particles in laboratories on Earth \cite{Sikivie:2020zpn} or through their imprint on astrophysical and cosmological observables \cite{Marsh:2015xka}.  
The most generic and well-studied interaction involves one axion and two photons; in the context of quantum field theory, this interaction is $\mathscr{L}_\mathrm{int} = - \tfrac{1}{4} \gagg a F_{\mu\nu} \tilde{F}^{\mu\nu}$ where $\gagg$ is the axion-photon coupling parameter, $a(x)$ is the axion field (its quantum excitations are axion particles), $F_{\mu\nu}(x)$ is the electromagnetic field strength tensor, and $\tilde{F}^{\mu\nu}(x) = \tfrac{1}{2} \epsilon^{\mu\nu\rho\sigma} F_{\rho\sigma}$.  

The axion-photon coupling $\gagg$ may be parametrized as $\gagg = - \mathcal{A} \alpha_\mathrm{em} / \pi f_a$ where $\alpha_\mathrm{em} \approx 1/137$ is the electromagnetic fine structure constant, $f_a$ is the energy scale of the spontaneously broken symmetry, and $\mathcal{A}$ is the electromagnetic anomaly coefficient.  
For a given theory, the anomaly is calculated as $\mathcal{A} = \sum_f \, Q_{\PQsc{},f} \, Q_{\mathrm{em},f}^2$ where $Q_{\PQsc{},f}$ is the Peccei-Quinn (\PQ{}) charge and $Q_{\mathrm{em},f}$ is the electric charge of $f$-species fermions \cite{Srednicki:1985xd}.  
Since the energy scale $f_a$ can vary by orders of magnitude from theory to theory \cite{Gendler:2023kjt}, it is difficult to identify target values of $\gagg \propto f_a^{-1}$ to motivate axion searches.  

Unlike $\gagg$, the anomaly $\mathcal{A}$ 
is generally expected to be an order-one rational number.  
For instance, minimal Grand Unified Theories (\GUT{}s) predict the anomaly to be an integer multiple of $\mathcal{A} = 4/3$ \cite{Srednicki:1985xd,Agrawal:2022lsp}.  
Furthermore, the quantization of electric charge (with down-type quarks having $Q_\mathrm{em} = 1/3$) requires $\mathcal{A}$ to be a multiple of $1/9$, implying that measurements of $\mathcal{A}$ test charge quantization \cite{Agrawal:2019lkr}.  
The strong theoretical expectation for $\mathcal{A} = O(1)$ motivates measurements that target the anomaly directly, such as anisotropic birefringence. 

When coupled to electromagnetism, the axion field becomes a birefringent medium 
\cite{Sikivie:1984yz,Huang:1985tt,Naculich:1987ci,Harvey:1988in,Carroll:1989vb,Carroll:1991zs,Harari:1992ea,Carroll:1998zi,Lue:1998mq}, \ie{} the polarization axis of linearly polarized light rotates as it propagates.  
For an observer looking outward into the axion field, the polarization rotation angle is calculated as 
\begin{align}\label{eq:alpha_def}
	\alpha(\nvec) = \frac{\gagg}{2} \int_{R(\nvec)} \dd X^\mu \, \partial_\mu a(X)\,,
\end{align}
where $R(\nvec)$ is a ray on the observer's past light cone $X^\mu$ in the direction $\nvec$ extending backward to the light's source.  
This phenomenon provides a way of searching out axions using precision observations of the cosmic microwave background (\CMB{}) polarization \cite{Komatsu:2022nvu} (or galaxies \cite{Yin:2024fez}).  
Depending on the configuration and evolution of the axion field, $a(x) = a(\xvec,t)$, a stronger or weaker signal may result.  
For models of axion dark matter or dark energy \cite{Finelli:2008jv,DeRocco:2018jwe,Obata:2018vvr,Fedderke:2019ajk,Fujita:2020aqt,Fujita:2020ecn,Obata:2021nql,Nakagawa:2021nme,Choi:2021aze,Nakatsuka:2022epj,Gasparotto:2022uqo,Naokawa:2024xhn,Zhang:2024dmi,Shao:2025jhp}, the axion field remains close to zero $\Delta a /f_a \ll 1$, and the fundamental theorem of calculus implies $\alpha(\nvec) = (\gagg/2) [ a(\text{observer}) - a(\text{source})]$, which is typically small, thereby making detection challenging.  

\begin{figure}[!t]
\includegraphics[width=0.48\textwidth]{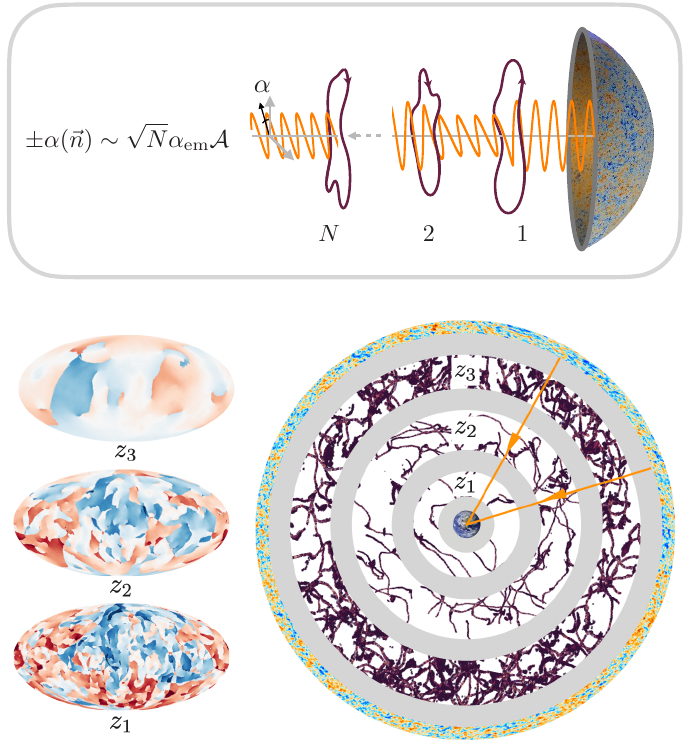}
\caption{\label{fig:illustration}
\textit{Top:}  An illustration of birefringence angle (polarization rotation) accumulated by passing through axion string loops. \textit{Bottom:}   The birefringence angle maps (left) as seen by observers at representative redshifts. The birefringence results from CMB photons passing through a network of axion strings. The maps are generated by ray-tracing through string-network simulations. For a roughly scaling network, the typical string correlation length grows with the horizon. Observers at low redshifts see more structure in the birefringence maps because of the smaller correlation length at early times and larger ones at later times.  
}
\end{figure}

However, if the axion field possesses topological defects known as cosmic strings (or domain walls \cite{Takahashi:2020tqv,Kitajima:2022jzz,Gonzalez:2022mcx,Ferreira:2023jbu,Lee:2025yvn}), then the change in the axion field can be much larger, and a stronger birefringence develops \cite{Naculich:1987ci,Harvey:1988in}. 
For a photon passing through a single loop of cosmic string, $\Delta a = \pm 2 \pi f_a$, and the birefringence is 
\begin{align}
	\alpha(\nvec) 
	= \mp \alpha_\mathrm{em} \mathcal{A} 
	\approx (\mp 0.0418^\circ) (\mathcal{A} / 0.1)
	\;.
\end{align}
With $N \gg 1$ such encounters, the birefringence accumulates as $\alpha \propto \sqrt{N}$ \cite{Agrawal:2019lkr,Jain:2021shf}, such that a telescope would observe a shadow of the string network; see \Fref{fig:illustration}.  
It is important to emphasize how the string-induced birefringence is insensitive to the symmetry breaking scale $f_a$, which cancels between the factors of $\gagg \propto f_a^{-1}$ and $\Delta a \propto f_a$, and consequently the observable is a direct probe of the electromagnetic anomaly coefficient $\mathcal{A}$ \cite{Agrawal:2019lkr}.  
Survival of the string network until after recombination requires the axions to be hyperlight, having mass $m_a < 3H_\mathrm{rec} \sim 10^{-28} \; \mathrm{eV}$; we assume $m_a < 3H_0 \sim 10^{-33} \; \mathrm{eV}$ such that the network survives until today. 

\paragraph*{Motivation~~---\!}
In prior work \cite{Jain:2021shf,Jain:2022jrp,Hagimoto:2023tqm,Hagimoto:2024sgw} (see also \Rrefs{Yin:2021kmx,Yin:2023vit}), some of the authors derived predictions for CMB birefringence arising from a cosmological network of axion string (and string-wall) defects.  
Those studies employed a framework called ``loop crossing model'' (\LCM{}) to calculate mock maps showing how the birefringence angle $\alpha(\nvec)$ is predicted to vary across the sky.  
It was found that a combination of SPT-pol and Planck measurements, which are compatible with zero anisotropic birefringence, places strong constraints on the presence of axion strings in the Universe today.

However, the earlier approach had a notable limitation.  
\LCM{}-derived predictions for two-point statistics, such as the angular power spectrum $C_\ell^{\alpha\alpha}$, exhibit a parametric degeneracy $C_\ell^{\alpha\alpha} \propto \mathcal{A}^2 \xi_0$ between a parameter controlling the strength of the axion-photon coupling $\gagg \propto \mathcal{A}$ and a parameter controlling the number of strings $N_\mathrm{str} \propto \xi_0$.  
The non-detection of anisotropic birefringence by SPT-pol led to $\mathcal{A}^2 \xi_0 < 3.7$ (at $95\%$ C.L.) \cite{Jain:2022jrp} but could not disentangle $\mathcal{A}$ and $\xi_0$. 
Since next-generation CMB telescopes expect to significantly improve on polarization measurements \cite{BICEPKeck:2024cmk,AtacamaCosmologyTelescope:2025blo,LiteBIRD:2025yfb}, possibly leading to a detection of birefringence \cite{Pogosian:2019jbt}, it is important to critically assess the accuracy of axion string birefringence predictions and to develop a strategy that breaks the parameter degeneracy. 

\paragraph*{Question~~---\!}
Numerical lattice simulations of global string network dynamics inform our understanding of string sizes and abundances in a cosmological axion string network \cite{Davis:1986xc,Harari:1987ht,Chang:1998tb,Yamaguchi:1998gx,Yamaguchi:1999dy,Hagmann:2000ja,Hiramatsu:2010yn,Hiramatsu:2010yu,Hiramatsu:2012gg,Kawasaki:2014sqa,Fleury:2015aca,Klaer:2017qhr,Klaer:2017ond,Gorghetto:2018myk,Kawasaki:2018bzv,Hindmarsh:2019csc,Gorghetto:2020qws,Buschmann:2021sdq,Pierobon:2023ozb,Saikawa:2024bta,Kim:2024wku,Gorghetto:2024vnp,Correia:2024cpk,Kim:2024dtq,Benabou:2024msj,Correia:2025nns}.  
If these simulations could be used to ``calibrate'' the phenomenological string network parameters of the \LCM{}, then a more accurate and robust prediction for CMB birefringence could be derived.  
Currently only a handful of groups around the world have the computing resources needed for these large-scale and high-resolution simulations.  
In this work we use the output of a single global string network lattice simulation \cite{Benabou:2024msj} that employed adaptive mesh refinement (\AMR{}) \cite{Drew:2019mzc,Buschmann:2024bfj}.  

\paragraph*{Overview~~---\!}
Our approach is briefly summarized as follows.  
We calculate the birefringence map $\alpha(\nvec)$ by propagating light rays through the single \AMR{} simulation and evaluating the integral in \Eref{eq:alpha_def}.  
We generate many such maps by constructing $100$ pseudo-realizations of the \AMR{} simulation volume (see below), and by averaging over this ensemble we calculate the angular power spectrum $C_\ell^{\alpha\alpha}$.  
There are several reasons why this $C_\ell^{\alpha\alpha}$ is not a direct prediction of \CMB{} birefringence: (1) the simulation was performed under the assumption of a radiation-dominated early universe cosmology, but birefringence develops after recombination in the dark matter- and dark energy-dominated eras; (2) the duration of the simulation is shorter than the time elapsed between recombination and today (\ie{}, ratio of final to initial scale factor); and (3) the \AMR{} simulation indicates a logarithmic growth in the density of strings (\ie{}, $\xi$), which is of order one during the simulation but which should be order-ten when extrapolated until today.  

To address these issues, we use the angular power spectrum $C_\ell^{\alpha\alpha}$ calculated from the \AMR{} simulation to calibrate the \LCM{}.  
This calibration entails fitting the \LCM{} parameters characterizing the density of strings and their lengths by requiring agreement in the angular power spectrum $C_\ell^{\alpha\alpha}$.  
After calibrating the \LCM{}, only the anomaly $\mathcal{A}$ remains as a free parameter.  
We calculate $C_\ell^{\alpha\alpha}$ for an \LCDM{} cosmology, allowing for a conservative order-one extrapolation uncertainty.  
Using measurements of anisotropic birefringence by several \CMB{} telescopes, we derive upper bounds on $|\mathcal{A}|$ that strongly constrain the axion's \UV{} embedding.  
Finally, we remark on the implications for \CMB{} isotropic birefringence. 

\paragraph*{\bf \AMR{} simulation data~~---\!}
The axion string network simulation that was performed by the authors of \Rref{Benabou:2024msj} employed \AMR{} techniques to resolve the string cores while also encompassing a large simulation volume.  
Although the adaptive mesh was needed to ensure a reliable characterization of the string dynamics, the coarsest resolution in the simulation is sufficient for our calculation of anisotropic birefringence which is restricted to angular scales that are much larger than the angular extent of a string core. 
Over one hundred snapshots at this coarse resolution were made available to us \cite{Malte_AMR}.
The first snapshot is at conformal time $\eta_1$ (defined as the time when $H(\eta_1) = f_a$), the last snapshot is at $\eta_{117} = 117 \eta_1$, and the spacing is $\Delta \eta = \eta_1$.  
Nine snapshots are missing, which we handle by interpolating between the previous and subsequent snapshots.  
Our analysis only uses the $101$ snapshots from $\eta_{17}$ to $\eta_{117}$, thereby dropping the earliest time steps to ensure that transient behavior has had time to settle out \cite{Benabou:2024msj}. 
Each snapshot consists of a cubic lattice with $1024^3$ sites and periodic boundary conditions.  
The lattice represents a range of comoving spatial coordinates from $(x,y,z) = (0,0,0)$ to $(L,L,L)$ where $L = 200 / a_1 H_1$.  
At each site, the value of the complex scalar field $\Phi$ and its time derivative are saved.
Away from the string cores $\Phi(\xvec,\eta) \approx (f_a / \sqrt{2}) \, \exp[\ii a(\xvec,\eta)/f_a]$, and along a closed path that encircles a segment of string, the axion field varies $a \to a \pm 2 \pi f_a$, where the sign depends on the relative orientation of the path and string.  

\paragraph*{Birefringence calculation~~---\!}
We use the \AMR{} simulation data to calculate anisotropic birefringence $\alpha(\nvec)$ using a spatio-temporal discretization of the integral in \Eref{eq:alpha_def}.  
We place the ``observer'' on the final snapshot $\eta = \eta_{117}$ at the center of the cubic simulation volume, $(x,y,z) = (L/2, L/2, L/2)$.  
The observer's past light cone intersects earlier and earlier snapshots on larger and larger 2-spheres until eventually reaching $\eta = \eta_{17}$ where the 2-sphere inscribes the cubic simulation volume (\ie{}, its diameter is $L$).  
We approximate the integral along a ray lying on the past light cone \eqref{eq:alpha_def} as a ``staircase'' that traverses a distance $L/200$ through each of the snapshots before rising up to the next snapshot.  
A detailed explanation of these methods is provided in \Sref{sec:AMR_methodology} of the \SM{}, where we also validate the staircase approximation against an alternative interpolation method.  

We employ the \Healpix{} software package \cite{Gorski:2004by} to decompose the observer's 2-sphere onto discrete pixels, each corresponding to a ray along the past light cone.  
We take $N_\mathrm{side}=512$ corresponding to $3145728$ pixels that each span a solid angle of approximately $0.013 \; \mathrm{deg}^2$. Along each ray we sample $\Phi(\xvec,\eta)$ along the staircase approximation of the past light cone, using a linear interpolation between lattice sites.  
We calculate the change in the axion field as $\Delta a_i / f_a = \arg[ \Phi_{i+1} \, \Phi_i^\ast]$, and approximate the integral in \Eref{eq:alpha_def} by summing the $\Delta a_i$.  

\paragraph*{Pseudo-realizations~~---\!}
We would like to have multiple simulations in order to study the statistics of string-induced birefringence, such as the mean and variance of the monopole, but only a single \AMR{} simulation is available.  
However, the procedure described above did not make use of the full \AMR{} simulation data since the observer's past light cone only reaches a sphere inscribed within the cubic simulation volume.  
The data in the corners of the cube were not used in the calculation.  
Furthermore, we note that the simulation was performed using spatially periodic boundary conditions.  
In lieu of additional \AMR{} simulations, we use the single \AMR{} simulation data to create pseudo-realizations.  
This is done by randomly shifting the observer's position within the simulation volume.  
Although the pseudo-realizations are not completely independent, we find this approach to be more than adequate for quantifying the cosmic variance.  

\begin{figure}[!t]
\includegraphics[width=0.48\textwidth]{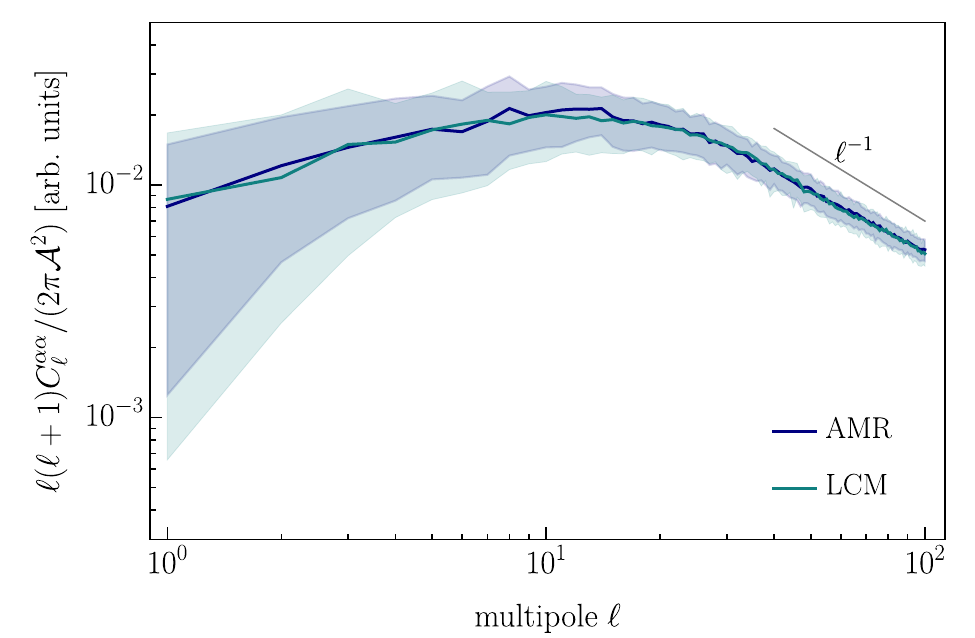}
\caption{\label{fig:calibration}
Calibrating the Loop Crossing Model (LCM) to the Adaptive Mesh Refinement (AMR) string network simulation.  We show the angular power spectrum $D_\ell^{\alpha\alpha} = \ell(\ell+1)C_\ell^{\alpha\alpha}/2\pi$ of anisotropic birefringence, which is calculated from $100$ pseudo-realizations of the AMR simulation and also from $100$ realizations of \LCM{} mock string networks. Both spectra scale with the electromagnetic anomaly coefficient as $D_\ell^{\alpha\alpha} \propto \mathcal{A}^2$.  Since the AMR simulation was performed in a radiation-era cosmology, we use the same background when generating the LCM mock networks.  We calibrate the LCM by fitting its parameters such that these two calculations of $D_\ell^{\alpha\alpha}$ agree. 
}
\end{figure}

\paragraph*{Anisotropic birefringence from \AMR{}~~---\!}
The procedure described above leads to $100$ pseudo-realizations of the birefringence map $\alpha(\nvec)$.  
We calculate the corresponding angular power spectra $C_\ell^{\alpha\alpha}$, and \Fref{fig:calibration} shows their mean and standard deviation.  
At large angular scales, the mean power spectrum is nearly scale invariant and the band reflects cosmic variance.  
The turnover at $\ell \sim 10$ is controlled by the characteristic angular scale of strings on the earliest snapshots as seen by an observer at $\eta = \eta_{117}$.  
The $\ell^{-1}$ tail is a consequence of the vanishing anisotropy on angular scales smaller than the characteristic string size \cite{Jain:2021shf}.  

\paragraph*{\bf The loop crossing model~~---\!}
The \LCM{} \cite{Jain:2021shf} may also be used to calculate the birefringence arising from a cosmological network of axion strings.  
It models the network as a collection of circular, planar, and static loops of string.  
If a photon passes through the disk bounded by a loop, it accumulates a birefringence $\Delta\alpha = \pm \mathcal{A} \alpha_\mathrm{em}$ depending on the relative orientation of the loop and the photon's path.  
Consequently, an \LCM{} birefringence map $\alpha(\nvec)$ resembles many overlapping elliptical regions.

The \LCM{} characterizes the string network using five parameters ($\xi_d$, $f_\mathrm{sub}$, $\zeta_\mathrm{min}$, $\zeta_\mathrm{max}$, and $\theta_\mathrm{cap}$).  
The dimensionless string density parameter $\xi_d$ enters the string network's energy density as $\rho_\mathrm{str} = \xi_d \mu d^{-2}$, where $d$ is the (monotonically growing) particle horizon, and the string tension $\mu$ drops out of the birefringence calculation. 
The dimensionless string length parameter $\zeta_d$ enters the string loop radius as $R = \zeta_d d$.  
The intuitive picture is that the network contains small and densely packed loops at early times when $d$ is small, and it contains large and sparsely packed loops at later times when $d$ is large.  
The \LCM{} does not describe the dynamical evolution of individual string loops, but rather captures the statistical properties of the network as a whole over cosmological timescales. 
In the mixed-loop model, a fraction \(f_{\rm sub}\) of the string length is distributed over \(\zeta_{\min} \leq \zeta \leq \zeta_{\max}\), and the remainder is at \(\zeta_{\max}\).
Finally, for loops that are nearby and have angular extent $\theta > \theta_\mathrm{cap}$, we suppress their contribution to the birefringence, taking a random $\Delta \alpha$ between $-\mathcal{A} \alpha_\mathrm{em}$ and $+\mathcal{A} \alpha_\mathrm{em}$, since the approximation $a = \pm 2 \pi f_a$ only holds for asymptotically far away loops.  
Additional details are provided in \Sref{sec:LCM} of the \SM{}.

\paragraph*{Calibrating the \LCM{}~~---\!}
To calibrate the \LCM{}, we adjust its parameters such that $C_\ell^{\alpha\alpha}$ derived from the \LCM{} mock string networks agrees with $C_\ell^{\alpha\alpha}$ derived from the \AMR{} simulation pseudo-realizations.  
We choose $\xi_d$ to match the logarithmically growing function of conformal time that was observed in the \AMR{} simulation.  
Since larger loops create more birefringence, $C_\ell^{\alpha\alpha}$ is insensitive to small $\zeta_\mathrm{min}$, and we establish convergence at $\zeta_\mathrm{min} = 0.003$.  
We scan \((f_{\rm sub},\zeta_{\max},\theta_{\rm cap})\), calculating $C_\ell^{\alpha\alpha}$ for 100 LCM realizations and comparing its mean with the mean and spread of $C_\ell^{\alpha\alpha}$ over $100$ AMR pseudo-realizations.   
The best fit point has $f_\mathrm{sub} = 0.6$, $\zeta_\mathrm{max} = 0.35$, and $\theta_\mathrm{cap} = 165^\circ$. 
We find that $\theta_\mathrm{cap}$ is weakly constrained, but this parameter does not impact $C_\ell^{\alpha\alpha}$ at the scales where bandpower measurements furnish constraints on $\mathcal{A}$.
The full calibration procedure and its score-weighted density are given in \Sref{sec:lcm_calibration}.

\Fref{fig:calibration} compares $C_\ell^{\alpha\alpha}$ 
derived from the \AMR{} simulation and the calibrated \LCM{}.  
At the level of the power spectrum (\ie{}, 2-point statistics) the agreement between these two approaches is excellent.  
In regard to shape, both methods reproduce the location and height of the peak around $\ell\sim 10$, as well as the $\ell^{-1}$ tail.  
More quantitatively, the mean $C_\ell^{\alpha\alpha}$ agree to within $10\%$ across the range of $\ell$ shown. 

\paragraph*{Extrapolation to late times~~---\!}
The \AMR{} simulation reveals the string dynamics for a window of time in the radiation era, soon after the network is formed.  
However, we seek to derive predictions for \CMB{} birefringence that accumulates between recombination and today.  
This mismatch is overcome with the calibrated \LCM{}.  

For the loop length parameter $\zeta_d$, we assume that the distribution of loop sizes remains unchanged between the early universe and today.  
In general this is a standard assumption for topological defect networks in the scaling regime \cite{Vilenkin:1982ks}, which is the case for the \AMR{} simulation at $\eta > \eta_{17}$ after transient effects have disappeared \cite{Benabou:2024msj}.  
For the string density parameter $\xi_d$, our fiducial calculation assumes that the slow logarithmic growth reported in \Rref{Benabou:2024msj} continues until today, which also follows from noting that the network is nearly in scaling.  
Given the enormous time intervals, $\xi_d$ grows by a factor of approximately $20$ between the end of the simulation and today.  
This estimate has a weak, logarithmic sensitivity to the mass scale of the radial mode, $m_r = \sqrt{2} f_a \sim 10^{12} \; \mathrm{GeV}$. 

We recognize that the extrapolation from early times until today is the dominant source of theoretical uncertainty in our calculation.  
To quantify this uncertainty, we use the phenomenological velocity one-scale model (\VOS{}) \cite{Martins:1995tg,Martins:1996jp,Martins:2000cs} to calculate the evolution of $\xi_d$ in an \LCDM{} cosmology.  
\VOS{} reveals that $\xi_d$ remains constant for a scaling string network in the radiation era, but it varies at radiation-matter equality when the universe becomes matter dominated and the cosmological expansion rate changes \cite{Coelho:2026oeg}. 
We find that $\xi_d$ can increase or decrease by as much as a factor of $2$ from the early universe till today, depending on the values of the \VOS{} parameters.  
Since $C_\ell^{\alpha\alpha} \propto \mathcal{A}^2 \xi_d$ \cite{Jain:2021shf}, we account for this extrapolation uncertainty (\EU{}) in $\xi_d$ as a scale-independent nuisance parameter $C_\ell^{\alpha\alpha} \to \kappa C_\ell^{\alpha\alpha}$ having a log-normal distribution with width $2$, centered at the fiducial $\xi_d$ extrapolation.  

\paragraph*{\bf Constraints on the coupling~~---\!}
The non-detection of anisotropic birefringence is reflected in measurements of $C_\ell^{\alpha\alpha}$ bandpowers that are consistent with zero \cite{POLARBEAR:2015ktq,Namikawa:2020ffr,SPT:2020cxx,Gruppuso:2020kfy,Bortolami:2022whx,BICEPKeck:2022kci}.  
For each telescope (and jointly), we calculate an upper limit on the electromagnetic anomaly $|\mathcal{A}| \leq \mathcal{A}_{95}$ (at 95\% C.L.) assuming a Gaussian likelihood for the reported bandpower measurements, zero covariance across bands, and uniform-per-mode averaging within each band.  

\begin{figure}[!t]
\includegraphics[width=0.48\textwidth]{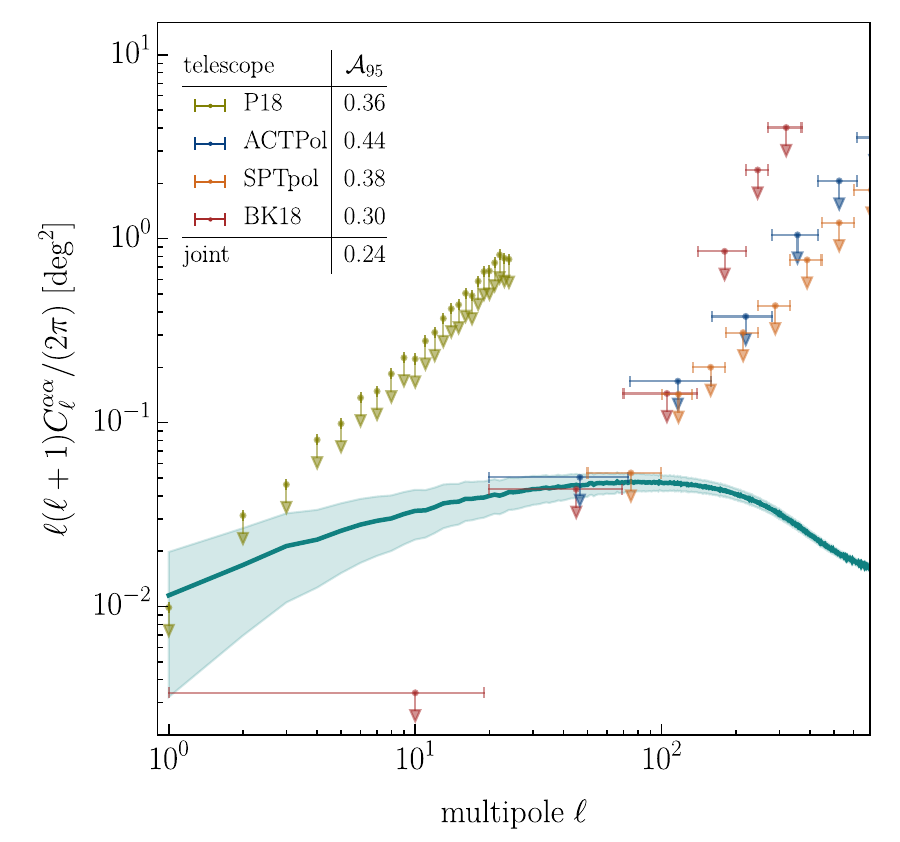}
\caption{\label{fig:anisotropic}
Predictions for \CMB{} anisotropic birefringence.  We show the angular power spectrum $C_\ell^{\alpha\alpha}$ of the birefringence rotation angle $\alpha(\nvec)$, which is calculated using the \LCM{} after calibrating to the \AMR{} simulation at early times and accounting for the factor of two extrapolation uncertainty (\EU{}) when evolving the string network to late times.  We set $\mathcal{A} = 0.24$  and more generally $C_\ell^{\alpha\alpha} \propto \mathcal{A}^2$. 
The teal curve and band show the mean and 16-84 percentile regions calculated from 500 realizations of the \LCM{} mock string network. Arrows indicate 95\% C.L. upper limits on $C_\ell^{\alpha\alpha}$ inferred from measurements of birefringence bandpowers by several \CMB{} telescopes, and the associated constraints on $\mathcal{A}$ appear in the table.  
}
\end{figure}

\paragraph*{\CMB{} anisotropic birefringence~~---\!}
\Fref{fig:anisotropic} shows the predicted \CMB{} anisotropic birefringence for the calibrated \LCM{}, accounting for the \EU{} that arises from extrapolating until late times.  
Similar to the \LCM{} curve appearing on \Fref{fig:calibration}, the angular power spectrum is approximately scale invariant at large angular scales and it falls as $\ell^{-1}$ at small angular scales.  
Here the turnover is shifted upward to $\ell \approx 100$, since the time interval between recombination and today ($a_0 / a_\mathrm{rec} \approx 1100$) is larger than the duration of the \AMR{} simulation ($a_{117} / a_{17} \approx 6.9$), which allows the network to contain smaller loops that support $C_\ell^{\alpha\alpha}$ at larger multipoles.  
We also show the 95\% C.L. upper limits inferred from measurements of $C_\ell^{\alpha\alpha}$ bandpowers by various \CMB{} telescopes.  
We find that the \EU{} degrades the limits by a factor of about $1.6$, and we interpret this result as a conservative treatment of the string network evolution and extrapolation. 

\paragraph*{CMB isotropic birefringence~~---\!}
Although measurements of anisotropic birefringence are consistent with zero, several recent studies \cite{Minami:2020odp,Diego-Palazuelos:2022dsq,Eskilt:2022wav,Eskilt:2022cff,Remazeilles:2025wzd,Diego-Palazuelos:2025dmh,Eskilt:2026imm} have identified a nonzero $EB$ cross-correlation in \CMB{} polarization data, which they interpret as evidence for an isotropic birefringence that is uniform across the sky.  
To align with conventions in the literature, we define the isotropic birefringence to be the opposite of the sky-average of our birefringence rotation angle $\beta = - \int \dd^2 \nvec \, \alpha(\nvec) / 4\pi$.  
Eskilt \& Komatsu (2022)~\cite{Eskilt:2022cff} derive $\beta = 0.342^{+ 0.094^\circ}_{- 0.091^\circ}$ using a joint analysis of WMAP 9-year and Planck PR4 data; 
Remazeilles (2025)~\cite{Remazeilles:2025wzd} derives $\beta = 0.32^\circ \pm 0.12^\circ$ using Planck PR4 data; 
Diego-Palazuelos \& Komatsu (2025)~\cite{Diego-Palazuelos:2025dmh} derive $\beta = 0.215^\circ \pm 0.074^\circ$ using ACT DR6 data; 
and Eskilt (2026)~\cite{Eskilt:2026imm} derives $\beta = 0.277^\circ \pm 0.057^\circ$ using a joint analysis of Planck PR4 and ACT DR6 data.  
See \Rref{Eskilt:2026imm} for a more complete enumeration of isotropic birefringence measurements and a discussion of how these analyses handle miscalibration uncertainty. 

Axion strings are not expected to generate an appreciable isotropic birefringence.  
A light ray is equally likely to pass by a string (or through a loop) with positive or negative winding number, causing positive and negative birefringence to accumulate with equal probability.  
Although the ensemble average of $\beta$ is zero, the value of $\beta$ in any single realization, such as our Universe, will be nonzero.  Consequently, one can ask whether the variance in $\beta$ is large enough to explain the reported detections of isotropic birefringence \cite{Jain:2022jrp,Ferreira:2023jbu}. 

\begin{figure}[!t]
\includegraphics[width=0.48\textwidth]{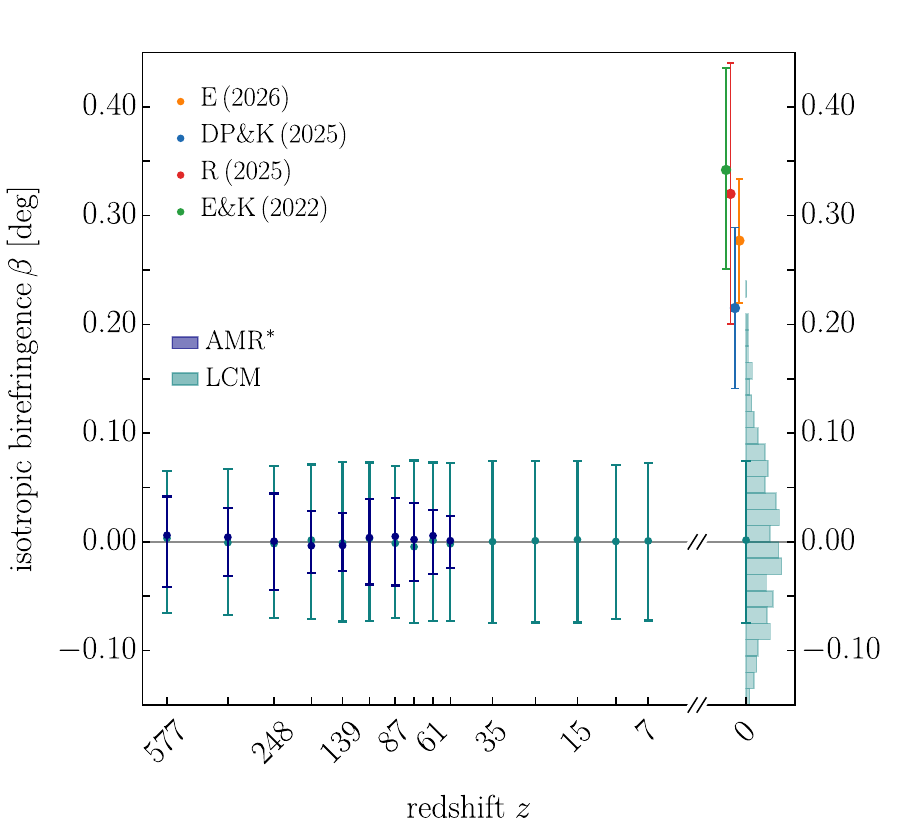}
\caption{\label{fig:isotropic}
Predictions for CMB isotropic birefringence.  The purple and teal dots and bars illustrate how the isotropic birefringence $\beta$ evolves with redshift $z$ between recombination ($z = 1100$) and today ($z = 0$).  The teal dots and bars represent mean and standard deviation over 1000 realizations of the calibrated \LCM{}, extrapolated to the late-time \LCDM{} cosmology.  We take $\mathcal{A} = 0.24$, which saturates the upper limit derived from constraints on anisotropic birefringence, and more generally $\beta \propto \mathcal{A}$.  The $z=0$ histogram represents our prediction for axion-string induced isotropic birefringence. Purple dots and bars were derived from $100$ pseudo-realizations of the AMR simulation. For comparison, we show four recently reported measurements of isotropic birefringence \cite{Eskilt:2022cff,Remazeilles:2025wzd,Diego-Palazuelos:2025dmh,Eskilt:2026imm}.
}
\end{figure}

Using the calibrated \LCM{} extrapolated to the late-time \LCDM{} cosmology, we calculate $\beta$ between recombination and today.  
In light of the constraints on $|\mathcal{A}|$ from anisotropic birefringence, we fix $\mathcal{A} = 0.24$, which saturates the joint 95\% C.L. upper limit.  
\Fref{fig:isotropic} shows the evolution of $\beta$ between recombination ($z_\mathrm{rec} = 1100$) and today ($z_0 = 0$), where the observable $\beta$ corresponds to the value today.  
For comparison, we show the rescaled evolution of $\beta$ derived from the \AMR{} simulation,  which runs over a shorter time interval so the purple points stop at $z \approx 52$. 
We observe that $\beta$ scatters around zero, as expected, since a positive or negative winding of the string network is equally likely.  
The standard deviation is approximately constant and equal to $\beta_\mathrm{rms} = 0.316^\circ \mathcal{A} = 0.075^\circ$.  
We find that the reported measurements lie in the tail of the predicted distribution. 
Among the three \LCM{} parameters being fit, the tails of the distribution are most sensitive to $\theta_\mathrm{cap}$.  
Taking a smaller value than our best fit would narrow the distribution, exacerbating the tension. 

\paragraph*{\bf Conclusions~~---\!}
In this work we adapted the output of an \AMR{} simulation of axion string network dynamics to calibrate the phenomenological \LCM{} framework and derive predictions for anisotropic and isotropic \CMB{} birefringence. 
This \AMR{}-assisted calibration breaks the $C_\ell^{\alpha\alpha} \propto \mathcal{A}^2 \xi_0$ degeneracy that limited predictivity in previous \LCM{}-based studies, and it allows us to infer an independent limit on the electromagnetic anomaly coefficient $\mathcal{A}$.  
A joint analysis of \CMB{} anisotropic birefringence measurements leads to $|\mathcal{A}| < 0.24$ at 95\% confidence, independent of the axion decay constant $f_a$.  
This is a powerful limit, since theories of fundamental physics in which axions couple to photons generally predict $\mathcal{A}$ to be an order-one rational number.  
E.g., this limit is inconsistent with a minimal \GUT{}, which predicts $\mathcal{A}$ to be an integer multiple of $4/3$, assuming hyperlight axions. 

Using the calibrated \LCM{}, we also make predictions for the isotropic birefringence $\beta$.  
Several recent studies derive $\beta \approx 0.3^\circ$ from measurements of $EB$ cross-correlation.  
We find that an axion string network predicts a range of values for $\beta$ that is symmetrically distributed around $\beta = 0$ with a standard deviation $\beta_\mathrm{rms} = 0.316^\circ \mathcal{A}$.  
In light of the tight upper bound from anisotropic birefringence $|\mathcal{A}| < 0.24$, we find that the axion string network is an unlikely explanation for the observed isotropic birefringence.  
For our fiducial model, the reported isotropic birefringence measurements lie in the tail of the predicted distribution. 

\paragraph*{Outlook~~---\!}
We find excellent agreement between the power spectra derived from the single \AMR{} simulation and the calibrated \LCM{}, although some calibration uncertainties remain.  
These could be reduced using an ensemble of existing fixed-lattice simulations.

The simulations used for our calibration are performed during the radiation era.  
A dominant uncertainty in connecting them to present-day observables arises from extrapolating the network evolution into the late-time \LCDM{} cosmology.  
Simulations spanning the radiation--matter transition could reduce this uncertainty.

We have focused on hyperlight axions having mass $m_a < 3H_0 \approx 10^{-33} \; \mathrm{eV}$, such that the string network survives until today.  
For $3H_0 < m_a < 3H_\mathrm{rec}$ the string network decays after recombination or forms a stable string-wall network, which impacts the limit on $\mathcal{A}$. 

Looking forward, axion-string-induced birefringence remains a compelling target for current and next-generation \CMB{} experiments, with the potential for stronger constraints on $\mathcal{A}$ or a discovery.

\acknowledgments
\paragraph*{\bf Acknowledgements~~---\!}We are grateful to Malte Buschmann for providing the \AMR{} simulation data files.  We thank Mark Hindmarsh, Asier Lopez, Carlos Martins, and Ken Olum for helpful responses to questions regarding string network scaling across different eras; and we thank Anna Coerver for insights on \CMB{} data analysis.  A.J.L. and M.V. are supported by the National Science Foundation under Grant No.~PHY-2412797.  M.A. and M.J. are supported in part by a U.S. Department of Energy grant DE-SC0010103.  M.W. was supported by Rice University's Summer Undergraduate Research Fellowship.  

\bibliographystyle{apsrev4-2.bst}
\bibliography{refs.bib}

\clearpage

\onecolumngrid
\makeatletter
\def\set@footnotewidth{\onecolumngrid}
\makeatother

\begin{center}
  \textbf{\large Supplemental Material for \\ \textit{CMB Birefringence from Axion String Networks Calibrated to an AMR Simulation}}\\[.2cm]
  \vspace{0.05in}
  {Mustafa A.~Amin, Mudit Jain, Andrew J.~Long, Aden Pugsley, Moira Venegas, and Magdalena Whelley}
\end{center}

\setcounter{equation}{0}
\setcounter{figure}{0}
\setcounter{table}{0}
\setcounter{secnumdepth}{2}
\setcounter{page}{1}
\makeatletter
\onecolumngrid
\renewcommand{\theequation}{S\arabic{equation}}
\renewcommand{\thefigure}{S\arabic{figure}}
\renewcommand{\thetable}{S\arabic{table}}

This Supplemental Material is organized as follows. Section~\ref{sec:AMR_methodology} describes the numerical methodology used to calculate birefringence from the Adaptive Mesh Refinement (\AMR{}) simulation, including the construction of the observer's past light cone and validation of the integration procedure. Section~\ref{sec:LCM} describes the Loop Crossing Model (\LCM{}) and the calculation of its birefringence observables. Section~\ref{sec:lcm_calibration} presents the calibration of the \LCM{} to the \AMR{} simulation and associated consistency checks. Section~\ref{sub:lcm_extrapolation} describes how the calibrated model is extrapolated from the radiation era to the late-time \LCDM{} cosmology, including the evolution of the string-network density. 
Finally, Sec.~\ref{sec:stats} describes the statistical procedure used to derive constraints on the electromagnetic anomaly coefficient $\mathcal{A}$ from measurements of anisotropic birefringence.

\section{Numerical methodology used to calculate \\ birefringence from the \AMR{} simulation}
\label{sec:AMR_methodology}

This section provides additional information about the \AMR{} simulation data, our staircase method for approximating integrals along an observer's past light cone, and a validation of the staircase method. 

\subsection{\AMR{} simulation data}
\label{sub:AMR_sim_data}

The numerical lattice simulation was performed in \Rref{Buschmann:2021sdq} by evolving in the \PQ{} complex field $\Phi(\vec x,\eta)$ in co-moving coordinates in a radiation dominated background.  
The simulation began before the \PQ{} phase transition, and the evolution captures spontaneously symmetry broken, followed by the formation and evolution of a network of global axion strings. 

The simulation volume consists of a cubic box with periodic boundary conditions and comoving side length $L = 200 / a_1 H_1$, discretized on a uniform lattice of $1024^3$ grid sites, where $a_1 \equiv a(\eta_1)$, is the scale factor at $\eta_1$, a reference conformal time such that $H_1 \equiv H(\eta_1) = f_a$.  
Although adaptive mesh refinement is employed in \Rref{Buschmann:2021sdq} to achieve high spatial resolution around the string cores throughout the evolution, the snapshots analyzed in this work correspond to the released data at the base grid resolution, without the additional refined sub-grids. 
This choice is justified by the fact that the CMB birefringence signal is determined by integrating the axion-field gradient over cosmological distances, the observable is therefore primarily sensitive to the large-scale configuration surrounding the strings rather than to the detailed internal structure of individual string cores.  

The system is evolved from an initial conformal time $\eta_{\rm in} = 0.1 \eta_1$ to $\eta_{\rm f} = 117 \eta_1$.  
At the initial time, $\eta_{\rm in}$, the simulated box contains $2000^3$ co-moving Hubble volumes, while by the final snapshot this number has decreased to roughly $\sim 6$ co-moving Hubble volume. 

We restrict our analysis to simulation snapshots spanning from $\eta_{17} = 17 \eta_1$ to $\eta_{117} = 117 \eta_1$ saved at intervals of $\Delta \eta = \eta_1$. 
The starting time is chosen to avoid transient effects associated with initial conditions.  
Seven snapshots within this range ($\eta/\eta_1$= 71, 74, 81, 85, 87, 90, 110) are missing from the released data, in the results presented here, we are using the nearest preceding snapshot to replace each missing snapshot.  

\subsection{Staircase method for integrating along the past light cone}
\label{sub:staircase}

\begin{figure}[!t]
    \centering
    \includegraphics[width=0.49\linewidth]{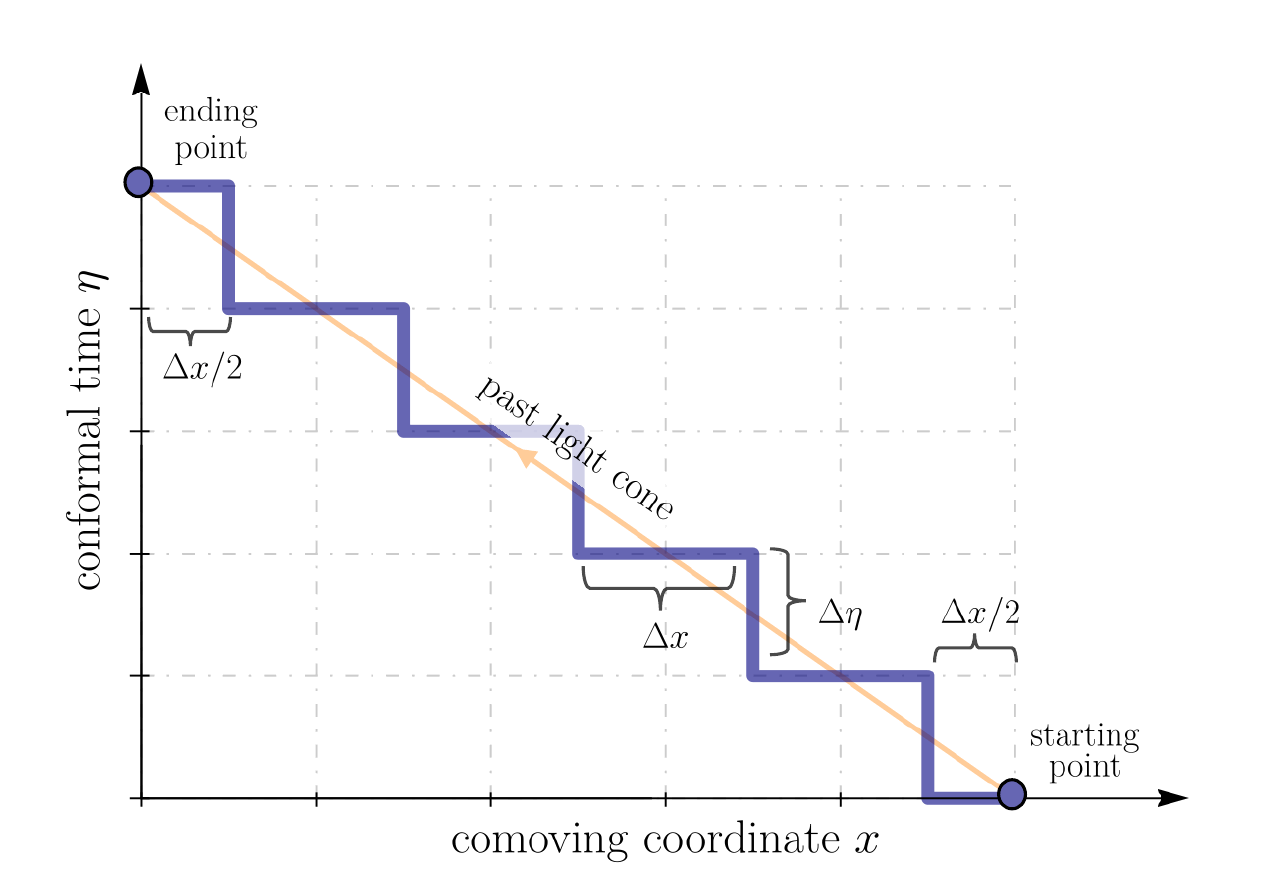}
    \includegraphics[width=0.49\linewidth]{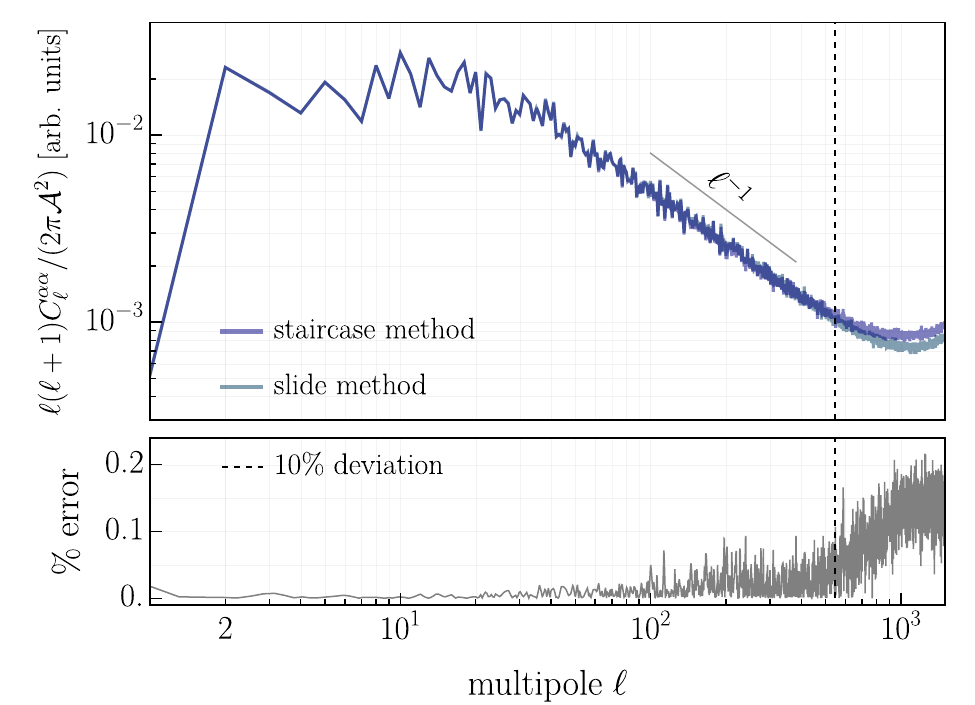}
        \label{fig:slide_check}
    \caption{Left: schematic of the staircase approximation to the past light cone. The thick dark blue trajectory approximates the true light cone (orange diagonal, $\dd\eta=|\dd x|$) with alternating fixed-$\eta$ (horizontal) and fixed-position (vertical) segments, stepping from the starting point at the earliest snapshot to the ending point at the observer. Right: validation of this scheme against the slide method (\Sref{sub:slide_check}) comparing the angular power spectrum for one realization at matched observed position and $N_\mathrm{side}=512$.}
    \label{fig:staircase}
\end{figure}

Photons travel on null geodesics satisfying $\dd\eta=
|\dd x|$, so that a light ray traces a $45^\circ$ path in the ($x,\eta$) plane, shown as the diagonal blue line in \Fref{fig:staircase}. The simulation output, however, is available only at discrete conformal times $\eta$, corresponding to individual snapshots separated by $\Delta\eta=\eta_1$ and on a discrete spatial lattice within each snapshot. As a result, the exact light-cone trajectory cannot be followed directly. We therefore replace the continuous light-cone with a discrete staircase trajectory, as illustrated in \Fref{fig:staircase}.

The staircase consists of a sequence of alternating horizontal and vertical segments. Along each horizontal segment, the photon propagates a fixed comoving distance at fixed conformal time $\eta$, using the field configuration of a single snapshot, the photon then advances $\Delta\eta$ to the next snapshot at fixed comoving position (a vertical segment). This sequence is repeated snapshot by snapshot until the photon reached the observer. The horizontal step size is $\Delta x$ for intermediate snapshots, and $\Delta x/2$ for the first and last snapshots along the ray.

In practice, this staircase construction is implemented by dividing the simulation volume into $N_{\rm{shells}}=101$ concentric spherical shells, with the outermost shell inscribed within the simulation box, \ie{}, at a comoving radius $r=L$ from the observer. Information in the corners of the box is excluded from the analysis. Each shell corresponds to a distinct simulation snapshot: the outermost shell corresponds to the earliest conformal time $\eta=17\eta_1$, and the innermost shell corresponds to the final snapshot $\eta=117\eta_1$. The width of the intermediate shells corresponds to $\Delta x=0.5L/(N_{\rm{shells}}-1)$ with $L=200\eta_1$.  

The light rays are initialized on the surface of the outer shell
using $\texttt{HEALPix}$ pixelization with $N_\mathrm{side}=512$, which sets the angular direction $\nvec$ and the angular resolution of the resulting birefringence map to be $\theta_{\rm{res}}=0.002\;\rm{rad}=0.115^\circ$. Each pixel corresponds to one line of sight that propagates radially inwards to the observer, stepping through the staircase trajectory described above. The integration along that trajectory produces a single values of $\alpha(\nvec)$. The number of sampling points $N_{\rm{points}}$ along each ray, between $r=L/2$ and $r=0$, is an input parameter that determines how many points are used to evaluate the field within each shell. We found that our results converge for $N_{\rm{points}}=1000$ leaving around 10 sample points per intermediate shell per ray.

Since a photon's position generally falls between lattice sites, we obtain the field there by linearly interpolating $\Phi$ on the lattice of the corresponding snapshot. At each sample point $j = 1, 2, \ldots, n$ along a ray, the interpolated value $\Phi_i\equiv\Phi(\vec x_i,\eta_i)$ is rescaled to unit modulus $\bar \Phi_i=\Phi_i/|\Phi_i|$, isolating its phase. The phase change between adjacent points then follows from the conjugate product,
\begin{equation}
    \Delta a_i/f_a = \arg \left[\bar\Phi_{i+1}\bar\Phi_i^*\right].
\end{equation}
Summing these increments over the full ray gives the net birefringence angle in direction $\nvec$
\begin{align}
    \alpha(\nvec) = \frac{\alpha_\mathrm{em} \mathcal{A}}{2\pi} \sum_i \Delta a_i / f_a 
    \;,
\end{align} 
which is in radians. 

To assess the statistical properties of the birefringence signal, it is necessary to average over multiple realizations of the string network.  
Since only one simulation box is available, and generating additional simulations is computationally expensive, we instead generate pseudo-realizations of the box by randomly shifting the observer's position within the simulation volume.  
Owing to the periodic boundary conditions, the field can be translated without introducing discontinuities, such that each observer effectively samples a distinct realization of the birefringence sky. 

For the computation of anisotropic birefringence, which is reported in \Fref{fig:anisotropic} of the article, we employ the staircase method described above, by constructing concentric shells centered on the observer at $\eta = 117 \eta_1$.  
Many pseudo-realizations of the \AMR{} simulation are generated by shifting the observer's position at random within the box. 
For the computation of isotropic birefringence, which is reported in \Fref{fig:isotropic} of the article, we adapt the staircase method to allow for a range of integration times.  We fix the initial time to be $\eta = 17 \eta_1$ and we vary the final time from $37 \eta_1$ to $117 \eta_1$.  
As the integration time increases, additional shells of fixed width are incorporated, and the shell corresponding to the initial time is displaced to progressively larger radii, ultimately reaching the edge of the simulation box when the integration time reaches $\eta=117\eta_1$. 
Each choice of final integration time thus yields a single realization of the birefringence angle $\alpha(\hat n)$.

\subsection{Validation of the staircase method}
\label{sub:slide_check}

As a check that the staircase method provides a reliable approximation to the past light cone, we implement a second method and compare the resulting angular power spectra.  
The second method, which we call the ``slide method,'' uses values of $\Phi$ and its time derivative $\Pi = \partial_\eta \Phi$ to interpolate between snapshots.  

For two consecutive snapshots in the simulation, $\eta_a$ and $\eta_b$ we construct the bracket $[\eta_a,\eta_b]$.  
The slide method uses both the field $\Phi$ and its conjugate momentum $\Pi\equiv\partial_\eta\Phi$ at $\eta_a$ and $\eta_b$.  
We build a cubic interpolation for intermediate $\eta_a<\eta<\eta_b$ as 
\begin{align}
    \Phi(\vec x,\eta)=c_0(\vec x) +c_1(\vec x)\eta+c_2(\vec x)\eta^2+c_3(\vec x)\eta^3 
    \;.
\end{align}
The four spatial-dependent coefficients are fixed by matching the sampled $\Phi$ and $\Pi$ to the bracket endpoints:
\begin{equation}
    \begin{pmatrix}
        \Phi_a\\
        \Pi_a \\
        \Phi_b\\
        \Pi_b
    \end{pmatrix} = \begin{pmatrix}
        1 & \eta_a & \eta_a^2 &\eta_a^3 \\
        0 & 1  & 2\eta_a & 3\eta_a^2 \\
        1 & \eta_b & \eta_b^2 &\eta_b^3 \\
        0 & 1  & 2\eta_b & 3\eta_b^2 \\
    \end{pmatrix}
    \begin{pmatrix}
        c_0\\
        c_1 \\
        c_2\\
        c_3
    \end{pmatrix}
    \;.
\end{equation}
We solve the system for the coefficients $c_i$ by inverting this matrix using numerical methods.  
Because the matrix only depends on $\eta_a$ and $\eta_b$, we only have to invert it once per bracket and reuse the inverse matrix for every sampled ray point falling within that bracket. 
In this slide method, the field is evaluated at each ray point's precise conformal time along the null geodesic, rather than remaining pinned to the snapshot value of $\eta$ as in the staircase approximation.

In the right panel in \Fref{fig:slide_check} we compare the resulting angular power spectrum $D_l^{\alpha\alpha}$ for one realization from each method, using the same observer position and $N_\mathrm{side}=512$ in both cases. The agreement between the two methods is excellent across the full multipole range: both curves track each other closely, both exhibit the same characteristic bend around  $\ell\sim 600$, reflecting the common HEALPix resolution limit shared by the two methods. Since the fractional error between them grows at exactly these resolution-limited scales, and since this has no bearing on the $\ell \sim16–200$ range, where the birefringence signal constraints are divided,  we  conclude that the staircase approximation introduces no significant error over the multipole range relevant to this work.

\section{The loop crossing model}
\label{sec:LCM}

In this section we explain how the loop crossing model (\LCM{}) describes the structure of a cosmological string network, we explain how the angular power spectrum and isotropic birefringence are calculated, and we explain the numerical methods and approximations that were used to generate realizations of mock birefringence maps.  

\subsection{String network structure in the \LCM{}}
\label{sub:LCM}

The loop-crossing model (\LCM{})~\cite{Jain:2021shf,Jain:2022jrp} is a phenomenological description of the string network on the observer's past light cone.  It represents the network by circular planar loops with random centres, orientations, and sizes.  The loops are treated as static during a photon crossing.  The purpose of the \LCM{} is to reproduce the accumulated crossing statistics of the network, rather than the microscopic evolution of individual string segments.

The energy density of the string network may be written as 
\begin{align}
    \rho_{\rm str}
    =
    \xi_d\,\mu\,d^{-2},
    \label{eq:lcm_rhostr_xid}
\end{align}
where \(\xi_d\) is the dimensionless string density parameter, \(\mu\) is the string tension, and $d$ is the particle horizon.  
Note that the particle horizon $d$ is related to the FRW scale factor $a$ and the conformal time $\eta$ by $d = a \eta$, throughout the entire cosmic history.  
Similarly the radius $R$ and comoving radius $R_\mathrm{com}$ of a string loop may be written as 
\begin{align}\label{eq:lcm_rcom_eta}
    R = \zeta d
    \qquad \text{and} \qquad 
    R_\mathrm{com}(\zeta) = R(\zeta) / a = \zeta\eta
    \;,
\end{align}
where $\zeta$ is the dimensionless string length parameter.  
Let \(p(\zeta)\,\dd\zeta\) denote the fraction of the total string length, equivalently the fraction of the string energy density, carried by loops with dimensionless radii in the interval \([\zeta,\zeta+d\zeta]\).  We normalize this distribution as
\begin{align}
    \int \dd\zeta\,p(\zeta)=1.
    \label{eq:lcm_loop_length_distribution_normalization}
\end{align}

Approximating the string length in a circular loop as \(2\pi R\), its energy is \(E_{\rm loop}=2\pi\mu\zeta d\). The corresponding physical and comoving number-density distributions are therefore
\begin{align}
    \frac{d n_{\rm phys}}{d\zeta}
    &=
    \frac{\rho_{\rm str}p(\zeta)}
         {2\pi\mu\zeta d}
    =
    \frac{\xi_d}{2\pi d^3}
    \frac{p(\zeta)}{\zeta} 
    \qquad \text{and} \qquad 
    \frac{d n_{\rm com}}{d\zeta}
    =
    a^3\frac{d n_{\rm phys}}{d\zeta}
    =
    \frac{\xi_d}{2\pi\eta^3}
    \frac{p(\zeta)}{\zeta} 
    \;.
    \label{eq:lcm_ncom_eta}
\end{align}
Thus, in conformal comoving variables, the scaling network depends on the cosmological background only through the relation between redshift and conformal time, together with the time dependence of \(\xi_d\).
For example, a ``monochromatic'' population of string loops, which all have the same radius at a given time, corresponds to 
\begin{align}
    p_{\rm mono}(\zeta) 
    = 
    \delta(\zeta-\zeta_0) 
    \qquad \text{and} \qquad 
    n_{\rm com}
    =
    \frac{\xi_d}{2\pi\zeta_0}\,\eta^{-3} 
    \;,
    \label{eq:lcm_ncom_monochromatic}
\end{align}
recovering the usual single-loop-size result.

Throughout this work, we adopt the mixed loop size variant of the \LCM{}.  
A fraction \(1-f_{\rm sub}\) of the total string length is carried by loops of size \(\zeta_{\max}\), and the remaining fraction \(f_{\rm sub}\) is distributed uniformly in \(\zeta\) between \(\zeta_{\min}\) and \(\zeta_{\max}\).  
In terms of the length distribution defined above,
\begin{align}
    p_{\rm mix}(\zeta)
    ={}&
    (1-f_{\rm sub})\,
    \delta(\zeta-\zeta_{\max}) + \frac{f_{\rm sub}}{\zeta_{\max}-\zeta_{\min}}\,
    \Theta(\zeta-\zeta_{\min})
    \Theta(\zeta_{\max}-\zeta).
    \label{eq:lcm_mixed_length_distribution}
\end{align}
The corresponding comoving loop-number distribution is
\begin{align}
    \frac{d n_{\rm com}}{d\zeta}
    ={}&
    \frac{\xi_d}{2\pi\eta^3}
    \left[
    \frac{1-f_{\rm sub}}{\zeta}\,
    \delta(\zeta-\zeta_{\max}) + \frac{1}{\zeta} \frac{f_{\rm sub}}
    {\zeta_{\max}-\zeta_{\min}}\,
    \Theta(\zeta-\zeta_{\min})
    \Theta(\zeta_{\max}-\zeta)
    \right].
    \label{eq:lcm_mixed_ncom_distribution}
\end{align}
Here the factor \(1/\zeta\) in the first term is, of course, distributionally equivalent to \(1/\zeta_{\max}\). Writing it in this form makes the common \(p(\zeta)/\zeta\) origin of both terms explicit.
The total comoving number density can be written as
\begin{align}
    n_{\rm com}
    =
    \frac{\xi_d}{2\pi\eta^3}\,
    {\cal I}_\zeta 
    \qquad \text{where} \qquad 
    {\cal I}_\zeta
    =
    \frac{1-f_{\rm sub}}{\zeta_{\max}}
    +
    \frac{f_{\rm sub}}
    {\zeta_{\max}-\zeta_{\min}}
    \ln\!\left(
    \frac{\zeta_{\max}}{\zeta_{\min}}
    \right) \;.
    \label{eq:lcm_mixed_loop_count_factor}
\end{align}
Thus the subloop component is uniform in \(\zeta\) when weighted by string length, whereas its loop-number distribution is proportional to \(1/\zeta\) and uniform in $\log\zeta$.  
Lowering \(\zeta_{\min}\) therefore increases the number of sampled loops only logarithmically.  At the same time, the angular contribution of sufficiently small loops is suppressed by their projected area, and the angular spectrum approaches a stable small-\(\zeta_{\min}\) limit.

\subsection{Birefringence from the \LCM{}}
\label{sub:LCM_birefringence}

Pixelated maps of the birefringence rotation angle $\alpha(\nvec)$ are generated using \Healpix{}.  
Loop centers are distributed at random across the 2-sphere of the sky.  
Each loop is characterized by a dimensionless loop length parameter $\zeta$, a conformal time $\eta$, and a random orientation, which together determine its angular extent.  
For instance, a loop with face-on orientation has apparent angular diameter 
\begin{align}
    \theta_{\rm loop}(\eta,\zeta)
    =
    2\tan^{-1}\!\left[
    \frac{R_{\rm com}(\eta,\zeta)}{s(\eta)}
    \right],
    \label{eq:lcm_loop_angular_diameter}
\end{align}
where $s(\eta)=\eta_{\rm obs}-\eta$ is the comoving distance from the observer to the loop center. 
A loop contributes to the birefringence map $\alpha(\nvec)$ when a photon ray intersects the projected loop disk.  
In \Healpix{}, we paint each loop's contribution $\Delta\alpha$ to the map $\alpha(\nvec)$ onto all the pixels pierced by the projected loop polygon.  

If a loop is sufficiently far away from the observer, then the axion field changes by a full $|\Delta a| = 2 \pi f_a$ along the photon's trajectory.  
Such loops contribute the maximal $\Delta\alpha = \pm \mathcal{A} \alpha_\mathrm{em}$ to the birefringence map.  
We select the sign $\pm$ at random for each loop with 50-50 probability.  
However, if the distance from the observer to the loop center is not large compared to the loop's radius, then $|\Delta a| < 2 \pi f_a$, and the birefringence contribution is less than maximal.  

To account for the reduced birefringence for nearby loops, we introduce the angular cap parameter, $\theta_\mathrm{cap}$.  
For each loop, if \(\theta_{\rm loop} < \theta_{\rm cap}\) then $\Delta \alpha$ is drawn at random from the 2-element set $\{- \mathcal{A} \alpha_{\rm em}, \, + \mathcal{A} \alpha_{\rm em} \}$, but if \(\theta_{\rm loop} > \theta_{\rm cap}\) then $\Delta \alpha$ is drawn at random uniformly from the interval \([-\mathcal{A}\alpha_{\rm em},+\mathcal{A}\alpha_{\rm em}]\). 

\subsection{Angular power spectrum and isotropic birefringence}
\label{sec:lcm_observables}

The \LCM{} lets us calculate a birefringence map \(\alpha(\nvec)\).  
We create $N_\mathrm{sim} \gg 1$ such realizations --- $\alpha^{(1)}(\nvec)$, $\alpha^{(2)}(\nvec)$, $\cdots$, $\alpha^{(N)}(\nvec)$ --- by drawing different random numbers for the loop locations, orientations, and sizes.  
For each realization $r \in \{1, 2, \cdots, N_\mathrm{sim} \}$ we use \Healpix{} to extract the moments of the spherical harmonic decomposition, 
\begin{align}
    \alpha^{(r)}(\nvec)
    =
    \sum_{\ell m}
    \alpha_{\ell m}^{(r)} \, Y_{\ell m}(\nvec) 
    \;,
\end{align}
where $m$ runs from $-\ell$ to $+\ell$. 
We calculate the angular power spectrum $C_\ell^{\alpha\alpha}$ as 
\begin{align}
    C_\ell^{\alpha\alpha}
    =
    \frac{1}{2\ell+1}
    \frac{1}{N_\mathrm{sim}} \sum_{r=1}^{N_\mathrm{sim}} 
    \sum_m
    \bigl( \alpha_{\ell m}^{(r)} \bigr) \bigl( \alpha_{\ell m}^{(r)} \bigr)^\ast 
    \;.
\end{align}
We calculate the isotropic birefringence for each realization $\beta^{(r)}$ as 
\begin{align}
    \beta^{(r)}
    =
    -\frac{1}{4\pi}
    \int \dd\Omega\,
    \alpha^{(r)}(\nvec) 
    \;.
    \label{eq:lcm_monopole_def}
\end{align}
In a \Healpix{} map, the integral and the division by $4\pi$ are implemented as the average over pixels.  
We calculate the mean and root-mean-square variation across realizations as 
\begin{align}\label{eq:app_beta_rms_estimator}
    \beta = \frac{1}{N_\mathrm{sim}} \sum_{r=1}^{N_\mathrm{sim}} \beta^{(r)} 
    \qquad \text{and} \qquad 
    \beta_\mathrm{rms} = \sqrt{ \frac{1}{N_\mathrm{sim}} \sum_{r=1}^{N_\mathrm{sim}} \bigl( \beta^{(r)} \bigr)^2 }
\end{align}
The spherical harmonic decomposition leads to the relations 
\begin{align}
    \beta^{(r)}
    =
    -\frac{\alpha_{00}^{(r)}}{\sqrt{4\pi}} 
    \qquad \text{and} \qquad 
    \beta_\mathrm{rms}^2 
    =
    \frac{C_0^{\alpha\alpha}}{4\pi} 
    \;.
    \label{eq:lcm_monopole_c0_relation}
\end{align}

\subsection{Semi-analytical derivation of isotropic birefringence}
\label{sub:isotropic_analytical}

The \LCM{} also admits a direct, map-independent Poisson calculation of \(\langle\beta^2\rangle\). This provides the theoretical monopole prediction used below and applies to any specified cosmological history. The direct calculation follows from Poisson statistics of the loop ensemble, without constructing a \Healpix{} map. For a loop with assigned birefringence amplitude \(a_{\rm loop}\) and projected solid angle \(\Omega_{\rm proj}\), the sky-average contribution is
\begin{align}
  \Delta\beta_{\rm loop}
  =
  -\,a_{\rm loop}\frac{\Omega_{\rm proj}}{4\pi}.
  \label{eq:app_single_loop_monopole}
\end{align}
The loop amplitudes have zero mean.  Consequently, the cross terms between distinct loops vanish after averaging over the Poisson ensemble:
\begin{align}
 \left\langle\beta^2\right\rangle
 =
 \int \dd N\,
 \left\langle
 a_{\rm loop}^2
 \left(\frac{\Omega_{\rm proj}}{4\pi}\right)^2
 \right\rangle_{\rm orient,\zeta}.
 \label{eq:app_monopole_poisson}
\end{align}
For ordinary loops, \(\langle a_{\rm loop}^2\rangle = (\mathcal A\alpha_{\rm em})^2\). For loops satisfying \(\theta_{\rm loop}\geq\theta_{\rm cap}\), the amplitude is instead drawn uniformly from \([-\mathcal A\alpha_{\rm em},+\mathcal A\alpha_{\rm em}]\), giving \(\langle a_{\rm loop}^2\rangle = (\mathcal A\alpha_{\rm em})^2/3\). 

For a loop whose center lies at comoving position \(s\hat{\bm r}\) and whose plane has unit normal \(\hat{\bm m}\), let \({\cal I}(\nvec;s,\hat{\bm r},\hat{\bm m},R_{\rm com})\) be the indicator that the ray in direction \(\nvec\) pierces the loop disk.  More explicitly, the ray intersects the loop plane at
\begin{align}
    t(\nvec)
    =
    s\,
    \frac{\hat{\bm m}\cdot\hat{\bm r}}
         {\hat{\bm m}\cdot\nvec},
\end{align}
and
\begin{align}
    {\cal I}
    &=
    \Theta\!\bigl(t(\nvec)\bigr)
    \Theta\!\left[
    R_{\rm com}^{2}
    -
    \left|
    t(\nvec)\nvec
    -
    s\hat{\bm r}
    \right|^{2}
    \right].
    \label{eq:app_loop_piercing_indicator}
\end{align}
The fractional area of the sky covered by the projected loop is therefore
\begin{align}
    \frac{\Omega_{\rm proj}}{4\pi}
    =
    \int\frac{d^{2}\nvec}{4\pi}\,
    {\cal I}(\nvec;
    s,\hat{\bm r},\hat{\bm m},R_{\rm com}).
    \label{eq:app_projected_sky_fraction}
\end{align}

Next, to make the Poisson measure explicit, define
\begin{align}
    q
    &=
    \ln\left(\frac{\eta_{\rm obs}}{\eta}\right),
    \qquad
    x(q)
    =
    \frac{\eta_{\rm obs}-\eta}{\eta},
    \qquad
    \varrho
    =
    \frac{\zeta}{x}
    =
    \frac{R_{\rm com}}{s},
    \label{eq:app_monopole_variables}
\end{align}
where \(s=\eta_{\rm obs}-\eta\).  The number of loops in a light-cone shell, with sizes in \(d\zeta\), is
\begin{align}
    dN
    &=
    4\pi s^2\,\eta\,dq\,
    \frac{d n_{\rm com}}{d\zeta}\,d\zeta = 2x^2(q)\,\xi_d(q)\,dq\,
    \frac{p(\zeta)}{\zeta}\,d\zeta.
    \label{eq:app_monopole_poisson_measure}
\end{align}
Here \(dq=-d\eta/\eta\) is taken positive toward earlier times, and we used \(d=a\eta\) together with Eq.~\eqref{eq:lcm_ncom_eta}. Thus this form applies to any FLRW background, with the cosmology entering through the conformal-time history and through \(\xi_d(\eta)\).

Let \(\mu_{\rm inc}\in[0,1]\) be the absolute cosine of the loop inclination. The orientation-averaged geometrical response of a loop with angular-size ratio \(\varrho\) is
\begin{align}
    {\cal G}(\varrho)
    =
    g_{\rm cap}(\varrho)
    \int_0^1 d\mu_{\rm inc}\,
    \left[
    \frac{\Omega_{\rm proj}(\varrho,\mu_{\rm inc})}
         {4\pi}
    \right]^2,
    \label{eq:app_monopole_geometry_response}
\end{align}
where \(\Omega_{\rm proj}(\varrho,\mu_{\rm inc})\) is the solid angle enclosed by the projected loop boundary and
\begin{align}
    g_{\rm cap}(\varrho)
    =
    \begin{cases}
        1,
        & 2\tan^{-1}\varrho<\theta_{\rm cap},\\[2pt]
        1/3,
        & 2\tan^{-1}\varrho\geq\theta_{\rm cap}.
    \end{cases}
    \label{eq:app_monopole_cap_response}
\end{align}
The factor \(1/3\) accounts for the variance of the uniform large-loop amplitude distribution. Numerically, we evaluate \(\Omega_{\rm proj}(\varrho,\mu_{\rm inc})\) by radially projecting a 256-sided polygonal approximation to the circular loop boundary onto the unit sphere and summing the solid angles of the resulting spherical triangles. This calculation is independent of the HEALPix pixelization used for the map realizations.

\begin{figure}[!t]
    \centering
    \includegraphics[width=0.75\textwidth]{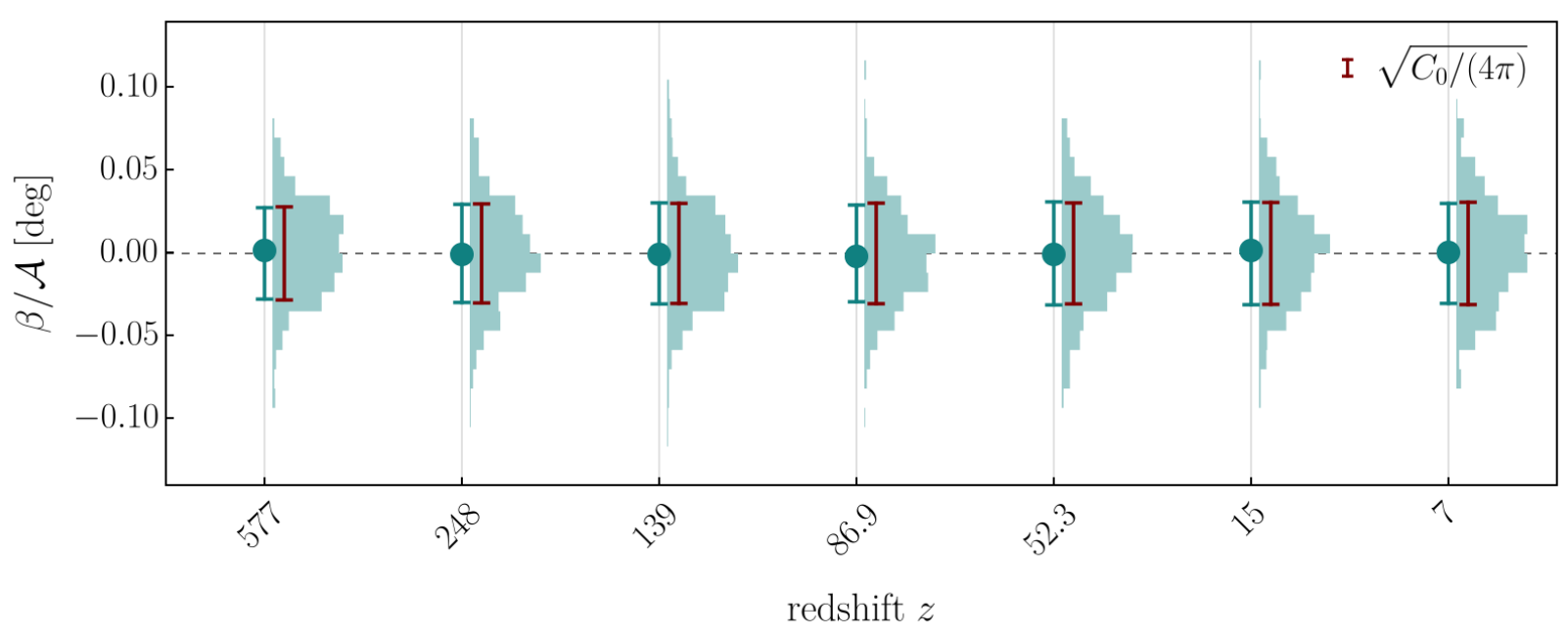}
    \caption{\label{fig:app_lcm_monopole_consistency}
    Internal \LCM{} monopole check. The histograms show the realization-to-realization distribution of the signed sky-averaged birefringence \(\beta\) at different final redshifts. For each distribution, we indicate the sample mean (dots) and the map RMS in \Eref{eq:app_beta_rms_estimator} (bars). The analytic \(\sqrt{C_0/(4\pi)}\) expectation is overlaid (in red) as a reference. This figure does not use \AMR{} data; it only checks the internal consistency of the \LCM{} monopole calculation.
    }
\end{figure}

For a general loop-length distribution \(p(\zeta)\), the size-averaged response is
\begin{align}
    {\cal R}_{\zeta}(x)
    =
    \int \dd\zeta\,
    \frac{p(\zeta)}{\zeta}\,
    {\cal G}\!\left(\frac{\zeta}{x}\right).
    \label{eq:app_monopole_general_response}
\end{align}
For the mixed-loop distribution in Eq.~\eqref{eq:lcm_mixed_length_distribution}, this becomes
\begin{align}
    {\cal R}_{\zeta}(x)
    ={}&
    \frac{1-f_{\rm sub}}{\zeta_{\max}}\,
    {\cal G}\!\left(\frac{\zeta_{\max}}{x}\right) + \frac{f_{\rm sub}}
    {\zeta_{\max}-\zeta_{\min}}\int_{\ln\zeta_{\min}}^{\ln\zeta_{\max}}
    d\ln\zeta\,
    {\cal G}\!\left(\frac{\zeta}{x}\right).
    \label{eq:app_monopole_mixed_response}
\end{align}
Combining Eqs.~\eqref{eq:app_monopole_poisson} and
\eqref{eq:app_monopole_poisson_measure} gives
\begin{align}
    \left\langle\beta^2\right\rangle
    =
    2(\mathcal A\alpha_{\rm em})^2
    \int_0^{q_{\rm ini}} \dd q\,
    x^2(q)\,\xi_d(q)\,
    {\cal R}_{\zeta}[x(q)],
    \label{eq:app_monopole_direct_c0}
\end{align}
where \(q_{\rm ini}=\ln(\eta_{\rm obs}/\eta_{\rm ini})\).  We evaluate this direct Poisson integral numerically, using the same \(\xi_d(\eta)\), mixed-loop distribution, and large-loop prescription as in the map calculation. Rotational invariance allows the loop centre to be fixed at the north pole; the remaining single-loop geometry is specified by its apparent radius \(\varrho=R_{\rm com}/s\) and inclination. We evaluate the orientation average in Eq.~\eqref{eq:app_monopole_geometry_response} using the polygonal solid-angle construction described above, and perform the remaining loop-size and light-cone integrals numerically.  No individual loops are drawn and no \Healpix{} map is constructed.  The result is therefore a direct, map-independent prediction for \(\sqrt{C_0/(4\pi)}\) for any chosen cosmological history.

For each \LCM{} map realization we calculate the signed sky average \(\beta^{(r)}\) directly from the \Healpix{} pixels, and the realization-to-realization RMS is given by \Eref{eq:app_beta_rms_estimator}.  We compare this map-based quantity with the direct Poisson prediction \(\sqrt{C_0^{\alpha\alpha}/(4\pi)} =\sqrt{\langle\beta^2\rangle}\), evaluated using \Eref{eq:app_monopole_direct_c0} and shown as the red markers in \Fref{fig:app_lcm_monopole_consistency}. This comparison tests both the finite-map implementation and the monopole extraction against a calculation that neither draws individual loops nor constructs a
\Healpix{} map.

\subsection{Numerical implementation}
\label{sec:numerical}

This following subsections describe the numerical implementation used to evaluate the calibrated \LCM{}.  We discuss the sampling of the past light cone, the treatment of finite map resolution, and internal checks of the Monte Carlo calculation. We discuss the sampling of the light cone, the treatment of unresolved projected loops, and comparisons between map-based quantities and their direct Poisson counterparts.

\subsection{Redshift binning}
\label{sec:numerical_redshift_binning}

Loop centers are sampled from an inhomogeneous Poisson process on the past light cone.  Writing its intensity in a time coordinate \(q\) as \(d\Lambda/dq\), the expected number of loops in a bin \(B_b\) is
\begin{align}
    \Lambda_b
    =
    \int_{B_b} \dd q\,
    \frac{d\Lambda}{dq}.
\end{align}
The binned implementation draws \(N_b\sim{\rm Poisson}(\Lambda_b)\) independently in each bin, and then samples the time of every loop from the normalized distribution \((d\Lambda/dq)/\Lambda_b\) within that bin.  We use \(n_{\rm step}=28\) bins in the production runs.

As a check, we also use \(n_{\rm step}=1\), with one Poisson draw from the full light-cone intensity followed by direct sampling from the full normalized time distribution.  The two procedures agree within Monte Carlo scatter.  This is expected because independent Poisson processes in disjoint bins superpose to the same inhomogeneous Poisson process, provided both the bin integrals and the conditional within-bin distributions are evaluated correctly.

\subsection{Subpixel loop painting}
\label{sec:numerical_subpixel}

A finite projected loop always overlaps at least one \Healpix{} pixel. However, the map rasterization represents a pixel by its central direction and initially paints a pixel only if that direction lies inside the projected loop polygon. A sufficiently small loop can therefore overlap several pixels while containing none of their
centres.

For a \Healpix{} map,
\begin{align}
    N_{\rm pix}
    &=
    12 \, N_\mathrm{side}^2 \;,
    \qquad
    \Omega_{\rm pix}
    =
    \frac{4\pi}{N_{\rm pix}},
    \label{eq:app_healpix_pixel_area}
\end{align}
where \(\Omega_{\rm pix}\) is the solid angle of one pixel. Whenever the polygon query returns no pixel centers, we calculate the projected solid angle \(\Omega_{\rm proj}\) directly from the vertices of the spherical polygon and define
\begin{align}
    N_{\rm eff}
    =
    \frac{\Omega_{\rm proj}}{\Omega_{\rm pix}}.
    \label{eq:app_neff}
\end{align}
Thus \(N_{\rm eff}\) is the effective number of equal-area \Healpix{} pixels covered by the loop.

We stochastically round \(N_{\rm eff}\) to an integer \(N_{\rm paint}\):
\begin{align}
    N_{\rm paint}
    =
    \lfloor N_{\rm eff}\rfloor
    +
    {\rm Bernoulli}\!\left(
    N_{\rm eff}-\lfloor N_{\rm eff}\rfloor
    \right).
    \label{eq:app_subpixel_rounding}
\end{align}
When \(N_{\rm paint}\geq1\), the pixel containing the loop centre is painted first.  Further pixels are selected from the frontier of its unpainted HEALPix neighbours, with equal probability, until the cluster contains \(N_{\rm paint}\) pixels.  Hence
\begin{align}
    \left\langle N_{\rm paint}\right\rangle
    =
    N_{\rm eff},
\end{align}
and the projected solid-angle weight is preserved in expectation. This construction is used only when the ordinary polygon query returns no pixel centres; it does not replace the resolved projected-loop geometry.

\section{Calibrating \LCM{} against \AMR{}}
\label{sec:lcm_calibration}

The mixed loop size \LCM{} has five free parameters: $\xi_d$, $f_\mathrm{sub}$, $\zeta_\mathrm{min}$, $\zeta_\mathrm{max}$, and $\theta_\mathrm{cap}$.  
In this section we describe how each of the parameter are determined, which we characterize as the ``calibration'' of the \LCM{}.  

\subsection{Calibrating $\xi_d$}
\label{sub:measured_xi}

Figure~2 of \Rref{Benabou:2024msj} presents a measurement of the string density parameter $\xi$ as a function of \(\log(m_r/H)\), where $m_r$ is the mass scale of the radial mode.  
They also provide an empirical fitting function: 
\begin{equation}\label{eq:benabou_xit_fit}
\begin{split}
    & \xi
    =
    c_{-2} \, L^{-2}
    +
    c_{-1} \, L^{-1}
    +
    c_0
    +
    c_1 L 
    \qquad \text{where} \qquad 
    L = \log(m_r/H) \\ 
    & \quad \text{and} \quad 
    c_{-2}=3.8,\qquad
    c_{-1}=0.13,\qquad
    c_0=-0.82,\qquad
    c_1=0.21\pm0.02 
    \;.
\end{split}
\end{equation}
The logarithmic growth in \Eref{eq:benabou_xit_fit} indicates a slow departure from scaling.

Note that $\xi \equiv \xi_t$ in \Rref{Benabou:2024msj} differs from the loop density parameter $\xi_d$ that we use to normalize the string energy density in the \LCM{}.  
These variables are related by $\rho_\mathrm{str} = \xi_t \, \mu \, t^{-2} = \xi_d \, \mu \, d^{-2}$ where $t$ is cosmic time and $d$ is particle horizon.  
The \AMR{} simulations were performed with the assumption of a radiation-dominated cosmological background, which corresponds to \(H=1/2t=1/a\eta=1/d\).  
With these changes of variable \Eref{eq:benabou_xit_fit} is equivalent to 
\begin{equation}\label{eq:xid_four_xit_rad}
\begin{split}
    & \xi_d
    =
    C_{-2} \, L^{-2}
    +
    C_{-1} \, L^{-1}
    +
    C_0
    +
    C_1 L 
    \qquad \text{where} \qquad 
    L = \log(m_r d) \\ 
    & \quad \text{and} \quad 
    C_{-2}=15.2,\qquad
    C_{-1}=0.52,\qquad
    C_0=-3.28,\qquad
    C_1=0.84\pm0.08
    \;.
\end{split}
\end{equation}
This formula represents our calibration of the loop density parameter $\xi_d$.  

\subsection{Calibrating $f_\mathrm{sub}$, $\zeta_\mathrm{min}$, $\zeta_\mathrm{max}$, and $\theta_\mathrm{cap}$}
\label{sub:calibrate_others}

To calibrate the other four \LCM{} parameters, we introduce a $\chi^2$ statistic that is minimized when $C_\ell^{\alpha\alpha}$ derived from \LCM{} and \AMR{} are equal.  
We use $100$ pseudo-realizations of the \AMR{} simulation, and $100$ \LCM{} realizations for each parameter set.  
We take $N_\mathrm{side} = 128$ for the \Healpix{} pixelization, and we compare $C_\ell^{\alpha\alpha}$ for $1 \leq \ell \leq 100$.  
We use this range because it contains both the low-multipole signal and the onset of the \(\ell^{-1}\) tail, while avoiding the resolution-sensitive high-\(\ell\) part of the maps.

The parameter $\zeta_\mathrm{min}$ sets the lower bound on the range of randomly drawn dimensionless loop radii $\zeta = R d = R_\mathrm{com} \eta$.  
Since larger loops contribute more to the birefringence, the power spectrum at $1 \leq \ell \leq 100$ is insensitive to $\zeta_\mathrm{min}$ if it takes values in the regime $\zeta_\mathrm{min} \ll \zeta_\mathrm{max}$.  
Thus after verifying convergence under decreasing \(\zeta_{\min}\), we fix
\begin{align}\label{eq:app_fixed_zeta_min}
    \zeta_{\min} = 0.003 
    \;.
\end{align}
This formula represents our calibration of the parameter $\zeta_\mathrm{min}$.  

The remaining three parameters are denoted collectively by $\vartheta = (f_{\rm sub}, \zeta_{\max}, \theta_{\rm cap})$.  
Let \(\overline D_{\ell,{\rm AMR}}\) be the mean of the $100$ \AMR{} pseudo-realizations, and let \(D_{\ell,{\rm AMR}}^{16}\) and \(D_{\ell,{\rm AMR}}^{84}\) be their 16th and 84th percentiles. We define
\begin{align}
    r_\ell(\vartheta)
    =
    \ln\!\left[
    \frac{D_{\ell,{\rm LCM}}^{\alpha\alpha}(\vartheta)}
         {\overline D_{\ell,{\rm AMR}}}
    \right] 
    \;, \qquad 
    s_{+,\ell}
    =
    \ln\!\left[
    \frac{D_{\ell,{\rm AMR}}^{84}}
         {\overline D_{\ell,{\rm AMR}}}
    \right]
    \;, \qquad \text{and} \qquad 
    s_{-,\ell}
    =
    \ln\!\left[
    \frac{\overline D_{\ell,{\rm AMR}}}
         {D_{\ell,{\rm AMR}}^{16}}
    \right] 
    \;,
    \label{eq:app_asymmetric_log_widths}
\end{align}
where $D_{\ell,{\rm LCM}}^{\alpha\alpha}(\vartheta)$ denotes the mean angular power spectrum, averaged over $100$ \LCM{} realizations.  
Our calibration statistic is the two-sided logarithmic score
\begin{align}
    \chi^2_{\rm asym}(\vartheta)
    =
    \sum_{\ell=1}^{100}
    \begin{cases}
    \left[r_\ell(\vartheta)/s_{+,\ell}\right]^2,
        & r_\ell(\vartheta)\geq0\;,\\[3pt]
    \left[r_\ell(\vartheta)/s_{-,\ell}\right]^2,
        & r_\ell(\vartheta)<0 \;.
    \end{cases}
    \label{eq:app_asymmetric_log_score}
\end{align}
If a set of \LCM{} parameters $\vartheta$ are found such that $\chi_\mathrm{asym}^2(\vartheta) = 0$ then the associated \LCM{} angular power spectrum is equal to the mean of the $100$ \AMR{} pseudo-realizations, \ie{} $D_{\ell,{\rm LCM}}^{\alpha\alpha}(\vartheta) = \overline D_{\ell,{\rm AMR}}$.  
A useful goodness-of-fit variable is 
\begin{align}
    {\cal E}_{\rm asym}(\vartheta)
    =
    \left(
    \frac{\chi^2_{\rm asym}(\vartheta)}{100}
    \right)^{1/2}
    \;,
    \label{eq:app_asymmetric_log_rms}
\end{align}
which also equals zero for an ideal fit.  
This definition retains the asymmetric realization-to-realization spread of the \AMR{} spectra.  
In particular, we do not impose an additional error floor. 

We scan over a cubic volume of the three-dimensional parameter space $\vartheta$.  
At each cell of the parameter space, we calculate $\chi_\mathrm{asym}^2(\vartheta)$ using \Eref{eq:app_asymmetric_log_score}.  
We find that the $\chi^2$ statistic is minimized at 
\begin{align}
    f_{\rm sub}=0.60 \;,
    \qquad
    \zeta_{\max}=0.35 \;,
    \qquad \text{and} \qquad 
    \theta_{\rm cap}=165^\circ 
    \;.
    \label{eq:app_best_fit_parameters}
\end{align}
These formulas represent our calibration of the parameters $f_\mathrm{sub}$, $\zeta_\mathrm{max}$, and $\theta_\mathrm{cap}$.  

In order to provide a visual representation of the goodness of fit across the parameter space, we perform the following calculation. 
The translated \AMR{} light cones are useful empirical realizations of the simulated volume, but they are not independent string-network simulations.  
Accordingly, \Eref{eq:app_asymmetric_log_score} is used as a calibration score rather than as a formal likelihood from which confidence intervals are inferred. 
For visualization, each parameter space cell \(i\) is assigned the normalized score weight $w_i$ and score-weighted density $\mathcal{P}_i$ according to 
\begin{align}
    w_i
    =
    \Delta V_i
    \exp\!\left[
    -\frac12\left(
    \chi^2_{{\rm asym},i}
    -
    \chi^2_{{\rm asym},\min}
    \right)
    \right]
    \qquad \text{and} \qquad 
    {\cal P}_i
    =
    \frac{w_i}{\sum_j w_j},
    \label{eq:app_score_weight}
\end{align}
where \(\Delta V_i\) is the parameter-space volume represented by the cell and $\chi^2_{{\rm asym},\min}$ is the smallest value of $\chi^2_{{\rm asym},i}$ across all cells. 
This construction accounts for the nonuniform spacings of the finite scan grid, but is not interpreted as a statistical posterior.

\begin{figure}[!t]
    \centering
    \includegraphics[width=0.8\textwidth]{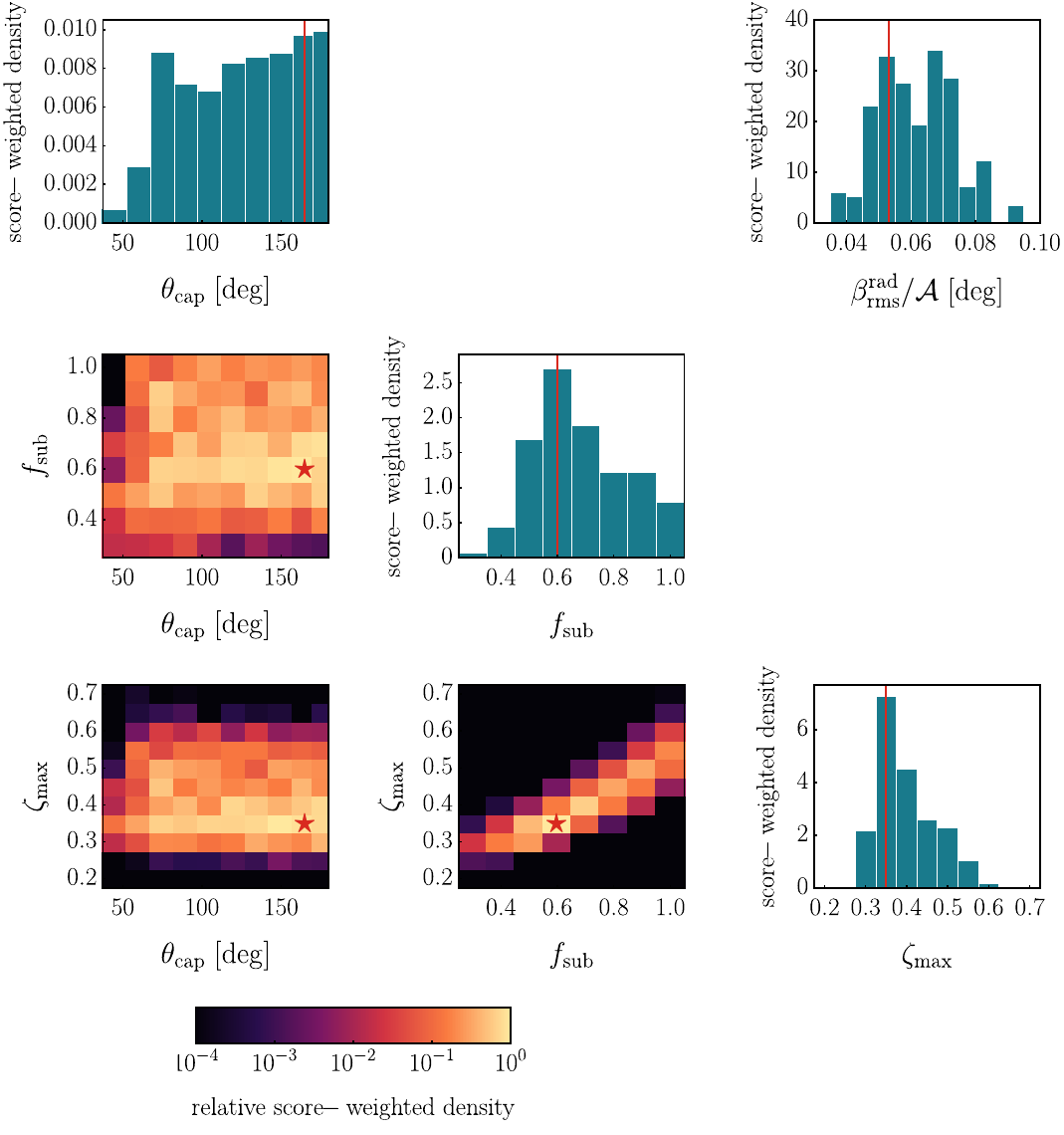}
    \caption{Score-weighted density over the mixed-loop calibration grid. The weights are defined in Eq.~\eqref{eq:app_score_weight}, using the asymmetric logarithmic score in Eq.~\eqref{eq:app_asymmetric_log_score}.  The red star marks the joint minimum of the score; the diagonal panels show the corresponding one-dimensional marginals. The lower-right panel gives the induced distribution of the radiation-era monopole RMS, \(\beta_{\rm rms}^{\rm rad}/\mathcal A\).}
    \label{fig:app_calibration_score_density}
\end{figure}

\Fref{fig:app_calibration_score_density} shows the score-weighted density $\mathcal{P}_i$, marginalized onto 1- and 2-dimensional slices of the 3-dimensional parameter space cube.  
In the $f_\mathrm{sub}$---$\zeta_\mathrm{max}$ plane, we observe a weak degeneracy direction with a tight constraint in the orthogonal direction.  
The weak degeneracy is likely associated with the net area covered by loops, since raising $f_\mathrm{sub}$ reduces the population of large loops at $\zeta = \zeta_\mathrm{max}$, whose abundance are proportional to $1-f_\mathrm{sub}$.  

The parameter $\theta_\mathrm{cap}$ is nearly unconstrained in the window $75^\circ \lesssim \theta_\mathrm{cap} \lesssim 170^\circ$.  
This parameter has the strongest effect on $C_\ell^{\alpha\alpha}$ at low multipoles, where the cosmic variance across the $100$ \AMR{} pseudo-realizations is large, allowing for a loose fit.  
Nevertheless, the power spectrum at higher $\ell$, where the \CMB{} bandpower measurements are most constraining, is insensitive to $\theta_\mathrm{cap}$ across this range of values.  
In particular the score-weighted 16--84-percentile envelope, the spread in \(D_\ell^{\alpha\alpha}\) is below approximately \(10\%\) for \(\ell\gtrsim20\). 
So the limit on $\mathcal{A}$ derived from bandpower measurements is robust despite the poor constraint on $\theta_\mathrm{cap}$. 
On the other hand, the value of $\theta_\mathrm{cap}$ has a stronger impact on the isotropic birefringence $\beta$, which is the $\ell = 0$ multipole. 
\Fref{fig:app_calibration_score_density} also shows the $\beta_\mathrm{rms}^\mathrm{rad}$, which is the root-mean-square of the isotropic birefringence over the $100$ \LCM{} realizations.  
We add ``rad'' to clarify that $\beta$ is begin calculated using the radiation era cosmology over the same time interval as the \AMR{} simulation, so these values of $\beta$ should not be interpreted as predictions for \CMB{} birefringence.  
Over the 3-dimensional parameter space volume, we find that $\beta_\mathrm{rms}^\mathrm{rad}$ is approximately 
\begin{align}
    \frac{\beta_{\rm rms}^{\rm rad}}{\mathcal A}
    =
    0.0615^{+0.0099}_{-0.0117}\ {\rm deg},
    \label{eq:app_rad_monopole_score_range}
\end{align}
where the quoted range is the score-weighted 16--84-percentile interval. Thus the 16--84 range spans approximately \(44\%\) of the median monopole RMS.

\subsection{Comparison of birefringence maps}
\label{sub:compare_maps}

\begin{figure}[!t]
   \centering
   \includegraphics[width=0.7\textwidth]{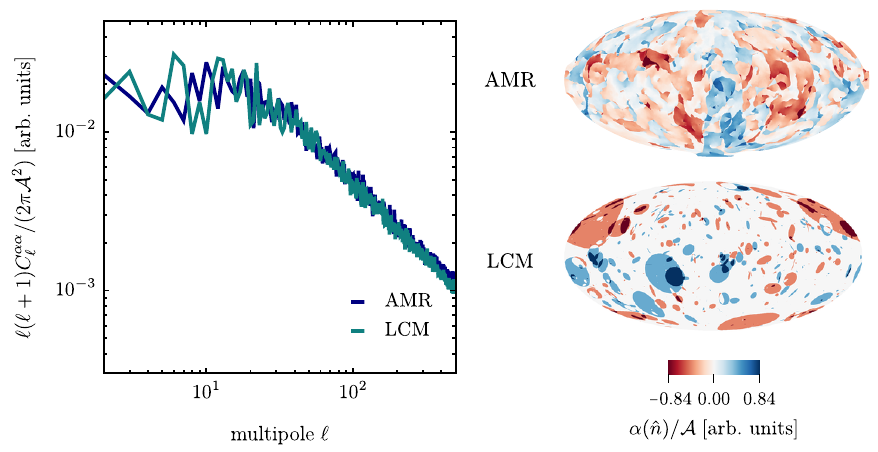}
   \caption{Representative single-realization comparison between the \AMR{} and calibrated \LCM{} birefringence fields. Right: Mollweide maps of \(\alpha(\nvec)/\mathcal A\) for the \AMR{} realization (upper panel) and an independent \LCM{} realization (lower panel). Left: the corresponding angular power spectra. The maps are not expected to agree realization by realization: the \LCM{} is calibrated to reproduce the statistical two-point signal, not necessarily higher-point structure of a particular \AMR{} map.
   } 
   \label{fig:ComparisonWithMaps}
 \end{figure}

The \LCM{} is a statistical description of the string network and is not expected to reproduce the morphology of an individual \AMR{} light cone.  \Fref{fig:ComparisonWithMaps} illustrates this point using representative single realizations.  Although the corresponding maps have visibly different structures, their power spectra are similar over the range relevant for the calibration.  The remaining differences reside in higher-point statistics, which are not fixed by the two-point calibration.

\subsection{Stability of the calibration across time}
\label{sec:cumulative_time_windows}

In order to derive predictions for \CMB{} birefringence, we must extrapolate the calibrated \LCM{} from the early universe until late times.  
By doing so, we assume that the \LCM{} accurately calculates the birefringence two-point statistics at times that are different from the calibration time.  
To validate this assumption, we compare \LCM{} and \AMR{} birefringence power spectra that accumulate over different time intervals.  
We use the best-fit calibrated \LCM{} \eqref{eq:app_best_fit_parameters}, and we calculate the power spectrum $C_\ell^{\alpha\alpha}$ of the birefringence $\alpha(\nvec)$ that accumulates between a fixed $\eta_i = 17 \eta_1$ and a variable $\eta_f$.  
Only the final observation time is changed; no parameters are refit. 

\begin{figure}[!t]
    \centering
    \includegraphics[width=\textwidth]{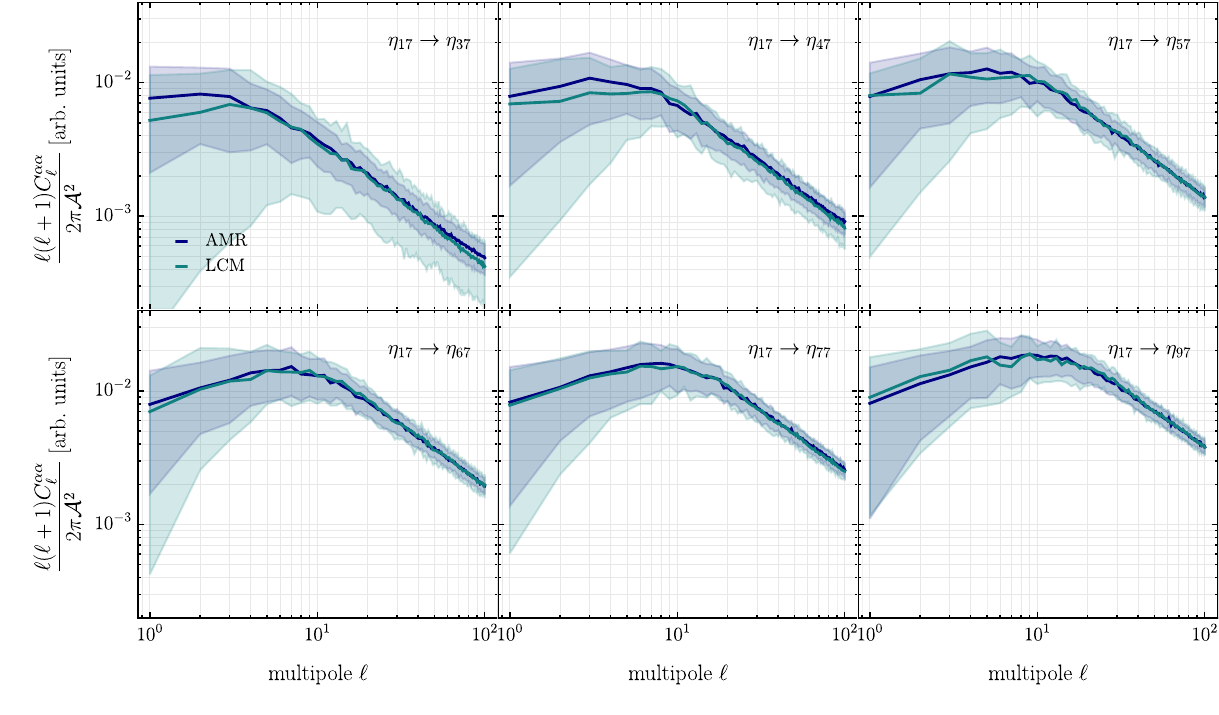}
    \caption{\label{fig:app_time_window_validation}
    Cumulative-window spectra for six values of \(\eta_{\rm f}\) from the calibrated mixed-loop \LCM{}. The parameters are fixed to \(f_{\rm sub}=0.60\), \(\zeta_{\min}=0.003\), \(\zeta_{\max}=0.35\), and \(\theta_{\rm cap}=165^\circ\) in every panel; no parameters are refit as \(\eta_f\) is varied.  Lines show the \AMR{} and \LCM{} mean spectra. The \AMR{} bands show one standard deviation across $100$ translated-light-cone pseudo-realizations, while the \LCM{} bands span the 16th--84th percentiles of 100 loop realizations. The full \(\eta_{17}\to\eta_{117}\) spectrum is shown in the main text and is therefore omitted here.
    }
\end{figure}

\Tref{tab:app_eta_window_scores} gives the asymmetric score \eqref{eq:app_asymmetric_log_rms} and the geometric mean spectrum ratio,
\begin{align}
    {\cal R}_{\rm geom}
    =
    \exp\!\left[
    \frac{1}{100}
    \sum_{\ell=1}^{100}
    r_\ell
    \right] 
    \;.
    \label{eq:app_geom_ratio}
\end{align}
\Fref{fig:app_time_window_validation} displays the spectra underlying \Tref{tab:app_eta_window_scores}. As \(\eta_{\rm f}\) is increased, the cumulative birefringence signal changes in both amplitude and angular shape.  The same fixed \LCM{} parameters follow this evolution across the six windows shown; the remaining differences and realization-to-realization scatter are quantified by \({\cal E}_{\rm asym}\) and \({\cal R}_{\rm geom}\) in the table.

\begin{table}[!h]
    \centering
    \caption{
    Cumulative-window comparison between the fixed calibrated \LCM{} and the \AMR{} spectra. The asymmetric log-residual score \({\cal E}_{\rm asym}\) and geometric mean ratio \({\cal R}_{\rm geom}\), defined in \Eref{eq:app_asymmetric_log_rms} and \Eref{eq:app_geom_ratio}, are evaluated over \(1\leq\ell\leq100\). Agreement improves as \({\cal E}_{\rm asym}\to0\) and \({\cal R}_{\rm geom}\to1\).
    }
    \label{tab:app_eta_window_scores}

    \begin{tabular}{
        c
        @{\hspace{0.35cm}} c
        @{\hspace{0.35cm}} c
        @{\hspace{0.35cm}} c
        @{\hspace{0.35cm}} c
        @{\hspace{0.35cm}} c
        @{\hspace{0.35cm}} c
        @{\hspace{0.35cm}} c
        @{\hspace{0.35cm}} c
        @{\hspace{0.35cm}} c
        @{\hspace{0.35cm}} c
    }
        \hline\hline
        \(\eta_f\)
        & 27 & 37 & 47 & 57 & 67 & 77 & 88 & 97 & 107 & 117 \\
        \hline
        \({\cal R}_{\rm geom}\)
        & 0.763 & 0.928 & 0.944 & 1.003 & 1.004
        & 0.979 & 0.998 & 1.006 & 0.985 & 0.988 \\
        \({\cal E}_{\rm asym}\)
        & 0.280 & 0.263 & 0.269 & 0.152 & 0.168
        & 0.211 & 0.144 & 0.182 & 0.203 & 0.175 \\
        \hline\hline
    \end{tabular}
\end{table}

The shortest windows contain relatively few loop crossings and consequently show the largest deviations.  For the longer windows, \({\cal R}_{\rm geom}\) remains close to unity.  Taken together, these comparisons test the reproducibility of the calibrated \LCM{} across the cumulative windows: the calibration fixed at \(\eta_f=117\) is applied unchanged at every other \(\eta_f\), including the shortest intervals.

\section{Extrapolating \LCM{} to late times}
\label{sub:lcm_extrapolation}

Axion string network simulations are generally performed in the radiation era of the early universe, but in order to derive predictions for \CMB{} birefringence we must characterize the string network at late times after recombination.  
A reasonable extrapolation can be performed by assuming that the network remains close to the scaling regime.  
In this section we describe how the extrapolation was performed, and we discuss its associated uncertainties.  

\subsection{\LCDM{} cosmology}
\label{sub:LCDM}

We model the cosmological expansion history as an \LCDM{} cosmology.  
We take the Hubble expansion rate $H$ to be 
\begin{equation}\label{eq:LCDM_cosmo}
\begin{split}
    & H = H_0 \, \sqrt{ \biggl( \frac{a}{a_0} \biggr)^{\!-4} \Omega_r + \biggl( \frac{a}{a_0} \biggr)^{\!-3} \Omega_m + \Omega_\Lambda} \\ 
    & \quad \text{with} \quad 
    \Omega_r = 9.17\times10^{-5},
    \quad
    \Omega_m = 0.315,
    \quad
    \Omega_\Lambda = 1-\Omega_m-\Omega_r,
    \quad \text{and} \quad 
    h=0.674 
    \;.
\end{split}
\end{equation}
This expression does not account for the decreasing effective number of relativistic degrees of freedom as the primordial plasma cools.  
We have checked that the effect on our extrapolation is at the sub-percent level.  
The conformal time $\eta$ and particle horizon $d$ are defined by
\begin{align}
    d = a\eta = a \int_0^a \! \frac{\dd a^\prime}{(a^\prime)^2 H(a^\prime)} 
    \;,
\end{align}
and we calculate the integral using numerical methods. 
Scale factor $a$ and cosmological redshift $z$ are related by $a = a_0 / (1+z)$, where $a_0$ is the scale factor today, $a_\mathrm{eq} = a_0 \Omega_r / \Omega_m \approx 2.91 \times 10^{-4} a_0$ is the scale factor at radiation-matter equality, and $z_\mathrm{rec} = 1100$ is the redshift at recombination. 

\subsection{String network in the scaling regime}
\label{sub:scaling}

A combination of numerical simulation and analytical arguments indicate that cosmological networks of local strings\footnote{
We are interested in the evolution of axion strings, which are global strings and not local strings.  In the following subsection, we account for the logarithmic departure from scaling that is observed in global string network simulations.} 
evolve toward an attractor solution, which is known as the scaling regime \cite{Vilenkin:2000jqa}.  
If the cosmology remains within a single cosmological era such that $a = a_\ast (t / t_\ast)^\lambda$, \eg{} radiation era ($\lambda = 1/2$) or matter era ($\lambda = 2/3$), then a scaling string network is characterized by constant values for the string density parameter $\xi_d$ and the string length parameter $\zeta$, which entered the comoving loop radius and the comoving loop density at \Erefs{eq:lcm_rcom_eta}{eq:lcm_ncom_eta}: 
\begin{align}\label{eq:lcm_projection_basic}
    R_{\rm com}(\eta,\zeta)
    =
    \zeta\eta
    \qquad \text{and} \qquad 
    \frac{d n_{\rm com}}{d\zeta}(\eta,\zeta)
    =
    \frac{\xi_d(\eta)}{2\pi\eta^3}\,
    \frac{p(\zeta)}{\zeta}
    \;.
\end{align}
However, if the cosmology evolves from one cosmological era into another (\eg{}, radiation era passes into matter era at radiation-matter equality), then all three of the $\xi$'s (and $\zeta$'s) cannot remain constant.  

In the following subsections we suppose that $\xi_d$ would remain constant for a scaling string network, even across different cosmological eras.  
We recognize that the axion string network simulations find slow logarithmic growth in $\xi_d$.  
We explain how we extrapolate the $\xi_d$ evolution from the early universe until today, in order to make predictions for \CMB{} birefringence.  
Finally we discuss the theoretical uncertainty associated with this extrapolation. 

\subsection{Extrapolating the string density parameter}
\label{sub:xi_extrapolation_cmb}

Our fiducial extrapolation assumes that the logarithmically growing $\xi_d$, which is observed in the \AMR{} simulation \eqref{eq:benabou_xit_fit}, remains applicable after radiation-matter equality provided that the time variable is expressed in terms of the particle horizon \eqref{eq:xid_four_xit_rad}.  
The string density parameter is therefore taken to be 
\begin{equation}\label{eq:xid_lcdm_extrapolated}
    \text{fiducial extrapolation:} \qquad 
    \xi_d(z)
    =
    C_{-2} \, L^{-2}
    +
    C_{-1} \, L^{-1}
    +
    C_0
    +
    C_1 L 
    \quad \text{where} \quad 
    L(z) = \log[m_r d(z)] 
    \;.
\end{equation}
A few example values are provided below: 
\begin{align}
    \xi_d \simeq
    C_1 \log\!\left(\frac{m_r}{10^{12}\,{\rm GeV}}\right) + \begin{cases}
    92.4, & z=1100,\\
    92.5, & z=1000,\\
    95.5, & z=100,\\
    98.3, & z=10,\\
    100.4, & z=1,\\
    100.9, & z=0.
    \end{cases}
    \label{eq:xid_lcdm_values}
\end{align}
Note that $\xi_d$ grows by a factor of approximately $25$ between the early universe ($L \approx 5-10$) and today ($z = 0$).  
However, the slow logarithmic growth leads to only a $\sim 9\%$ change recombination and today.

For numerical results throughout the article we take $m_r = 10^{12} \, \mathrm{GeV}$ and $f_a = m_r / \sqrt{2}$.  
These values only affect the birefringence calculation through the $\xi_d$ extrapolation, which is a weak logarithmic sensitivity.  

\begin{figure}[!t]
    \includegraphics[width=1\textwidth]
    {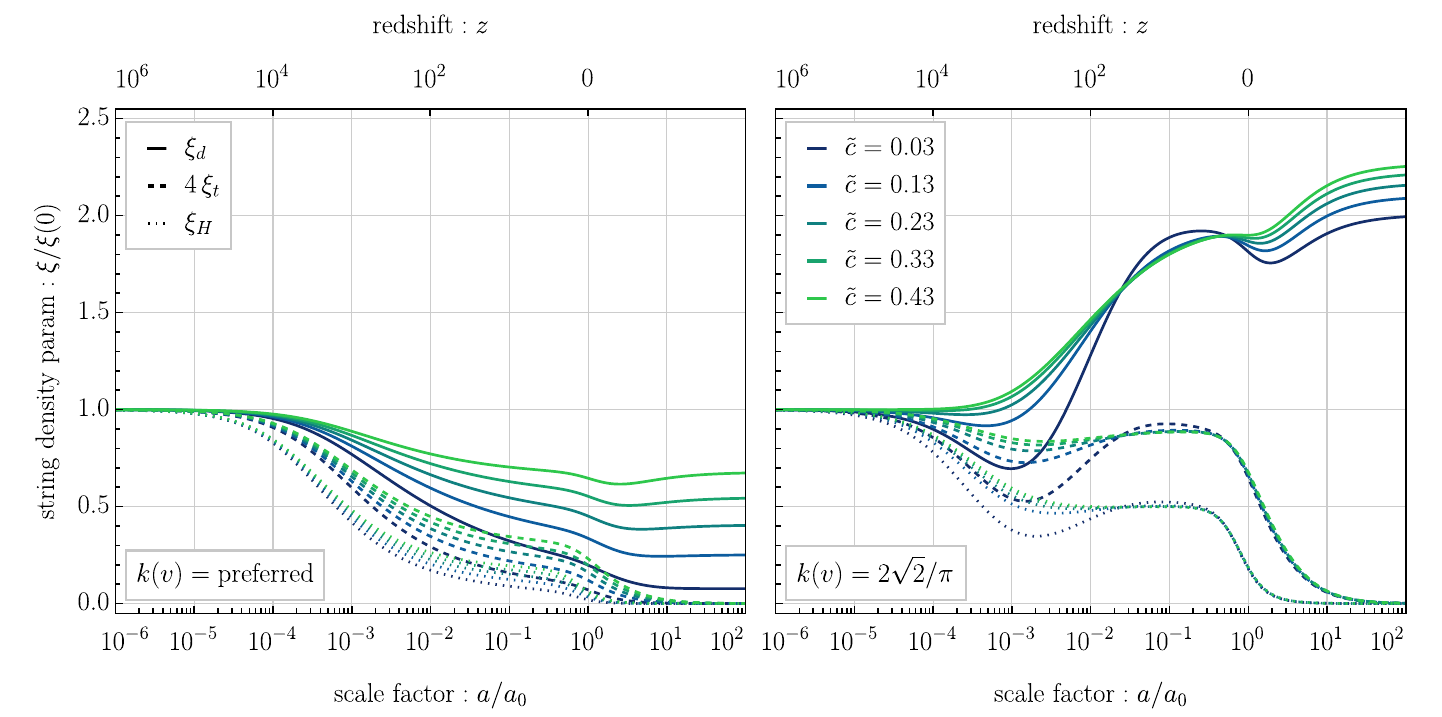}
    \caption{\label{fig:VOS} 
    Evolution of the loop abundance parameters \(\xi\) in \LCDM{} cosmology.  We plot solutions of the \VOS{} equations for several choices of the loop chopping efficiency $\tilde{c}$ and the momentum parameter $k(v)$.  As we vary the parameters away from their preferred values \eqref{eq:VOS_preferred}, the change in $\xi_d$ between radiation-matter equality (\(a/a_0 \approx 2.91 \times 10^{-4}\)) and today (\(a/a_0 = 1\)) can be as large as a factor of $2$.
    }
\end{figure}

\subsection{Extrapolation uncertainty}
\label{sub:extrapolation_uncertainty}

The fiducial extrapolation of $\xi_d$ in \Eref{eq:xid_lcdm_extrapolated} is motivated by the observation that $\xi_d$ remains constant for a scaling string network within a single cosmological era.  
However, to connect the calibration at early times to the observations of birefringence at late times, we need to extrapolate from the radiation era into the late-time \LCDM{} cosmology.  
We are not aware of numerical simulations studying the evolution of a scaling string network across cosmological eras.  
Here we draw upon a semi-analytical model of scaling string networks to infer the evolution of $\xi_d$ and quantify the theoretical uncertainty in our extrapolation.  

There exists an analytical and phenomenological description of scaling string networks that allows one to calculate the evolution of \(\xi_d\) across cosmological eras. 
This is known as the velocity one-scale (\VOS{}) model \cite{Martins:1995tg,Martins:1996jp,Martins:2000cs}, and the authors of \Rrefs{Martins:2018dqg,Coelho:2026oeg} have used \VOS{} to study axion string networks.  
The \VOS{} model has two dynamical variables: the string's root mean square velocity $v(t)$ and the characteristic length scale $L(t)$.  
Their evolution are governed by 
\begin{subequations}
\begin{align}
    2 \frac{\dd L}{\dd t} & = 2 H L \bigl( 1 + v^2 \bigr) + \tilde{c} v \\ 
    \frac{\dd v}{\dd t} & = \bigl( 1 - v^2 \bigr) \biggl[ \frac{k(v)}{L} - 2 H v \biggr] 
    \;,
\end{align}
\end{subequations}
where $H(t)$ is the Hubble parameter.  
In transcribing these equations from \Rref{Coelho:2026oeg} we have dropped the Aharonov-Bohm scattering term and the tension-varying term, and we have checked that they have a negligible impact on the late-time evolution. 
The two parameters are the loop chopping efficiency $\tilde{c}$ and the momentum parameter $k(v)$, which is a function of the string speed $v$.  
The authors of \Rref{Coelho:2026oeg} provide 
\begin{align}\label{eq:VOS_preferred}
    \text{preferred values:} \qquad 
    \tilde{c} = 0.23 
    \qquad \text{and} \qquad 
    k(v) = \frac{2 \sqrt{2}}{\pi} \bigl( 1-v^2 \bigr) \bigl( 1 + 2 \sqrt{2} v^3 \bigr) \frac{1 - 8 v^6}{1 + 8 v^6} 
    \;,
\end{align}
and they characterize these as the values that are preferred by numerical simulation. 

We solve the \VOS{} equations in an \LCDM{} cosmology, providing an initial condition such that the network is in scaling initially.  
The evolution of $\xi$ is shown in \Fref{fig:VOS}.
These variables are related by $\rho_\mathrm{str} = \xi_H \, \mu \, H^2 = \xi_t \, \mu \, t^{-2} = \xi_d \, \mu \, d^{-2}$ where $H$ is the Hubble parameter, $t$ is cosmic time, and $d$ is particle horizon.  
We confirm that a scaling string network has constant \(\xi_H = 4 \xi_t = \xi_d\) throughout the radiation era.  
After the radiation era is ended, we find that all three \(\xi\)'s vary by an order-one factor between radiation-matter equality and today.  
Moreover different choices of \(\tilde{c}\) and \(k(v)\) lead to different predictions for the evolution of \(\xi_d\).  
For the preferred parameters, $\xi_d$ decreases by a factor of $0.4$ between recombination and today, whereas for $k(v) = 2\sqrt{2}/\pi$ it instead increases by a factor of $1.6$.  
We interpret this variation as an inherent uncertainty in our extrapolation of the \(\xi_d\) evolution to late times. 

\begin{figure}[!t]
  \centering
  \includegraphics[width=0.6\textwidth]{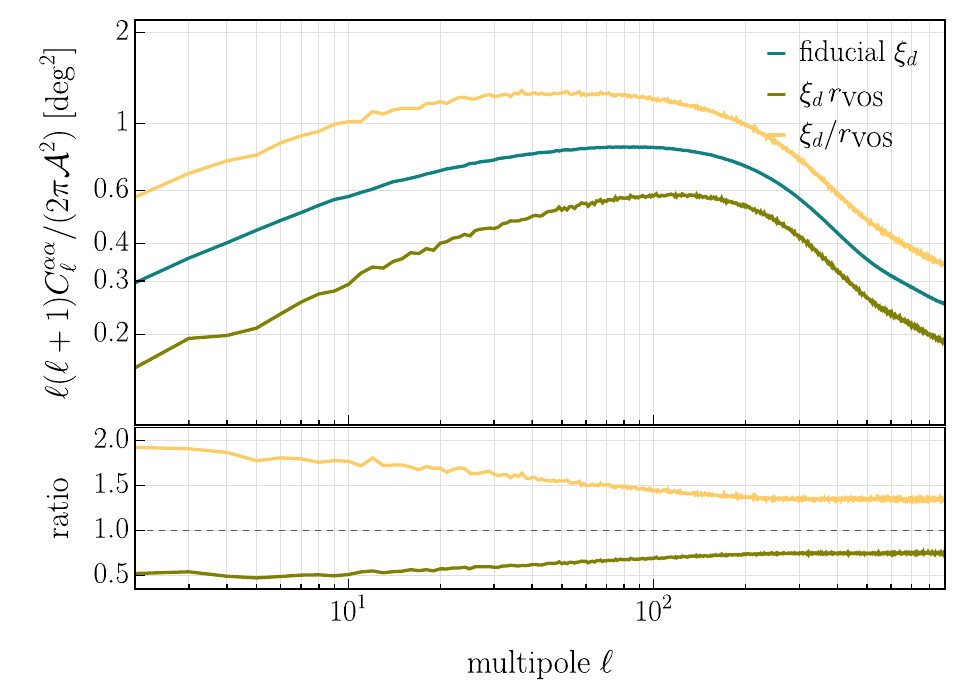}
  \caption{\label{fig:app_vos_xid_uncertainty}
  Effect of alternative \(\xi_d\) extrapolations on the calibrated \LCDM{} \LCM{} spectrum.  The blue curve is the fiducial log-growing \(\xi_d\) history.  The green curve multiplies the fiducial history by the \VOS{} ratio \(r_{\rm VOS}\), while the red curve uses the inverse trend \(1/r_{\rm VOS}\).  The modified histories primarily shift the normalization of the spectrum, with a mild low-\(\ell\) tilt.
  }
\end{figure}

To estimate how this extrapolation uncertainty impacts the anisotropic birefringence, we perform an empirical shape test.  
Using the \VOS{} solution for the preferred parameters, appearing in the left panel of \Fref{fig:VOS}, we define \(r_{\rm VOS}(a) = \xi_d(a)/\xi_d(0)\).  
We then use the calibrated \LCM{} to calculate the birefringence power spectrum $C_\ell^{\alpha\alpha}$ with the following three extrapolations of the string density parameter: 
\begin{align}\label{eq:xid_vos_variations}
    \xi_d^{\rm fid}(a) \;,
    \qquad
    \xi_d^{\rm fid}(a)\,r_{\rm VOS}(a) \;,
    \qquad \text{and} \qquad 
    \frac{\xi_d^{\rm fid}(a)}{r_{\rm VOS}(a)}
    \;,
\end{align}
where $\xi_d^\mathrm{fid}(a)$ is the fiducial extrapolation \eqref{eq:xid_lcdm_extrapolated}.  
The second extrapolation follows the trend of the \VOS{} curve, while the third gives the corresponding inverse trend in multiplicative, or log, space.  These two extrapolations are not intended as alternative best-fit network models, but as stress tests of how an order-one change in the late-time \(\xi_d\) evolution affects the \LCM{} angular spectrum.  The resulting spectra are shown in \Fref{fig:app_vos_xid_uncertainty}.  The dominant effect is an overall normalization shift, which either increases or decrease $C_\ell^{\alpha\alpha}$ by a factor of $\lesssim 2$.  There is also a mild tilt, most visible at low multipoles, but the template shape remains stable.  

We fold the extrapolation uncertainty into our constraints on \(\mathcal{A}\) using the statistical method explained in \Sref{sec:stats}.  
We introduce a nuisance parameter $\kappa$ as a multiplicative factor on the power spectrum template.  
\Fref{fig:VOS} motivates us to draw $\kappa$ from a log-normal distribution with width $2$, and \Fref{fig:app_vos_xid_uncertainty} motivates us to use an $\ell$-independent rescaling.  

\subsection{Evolution of isotropic birefringence}
\label{sub:isotropic_evolution}

Using both the \AMR{} and \LCM{} methods, we study the evolution of isotropic birefringence after recombination, and compare them on \Fref{fig:isotropic}.  
In this section we explain how that comparison was made, and we discuss how it serves as a test of the calibration.  

For the \AMR{} simulation we calculate $\beta$ over different time intervals starting at $\eta_i = 17 \eta_1$ and ending at a variable $\eta_f$ that ranges from $37 \eta_1$ to $117 \eta_1$.  
Each pseudo-realization gives a different random $\beta_f^{(r)}$ at each $\eta_f$, and we calculate $\beta_f$ by averaging over the pseudo-realizations.  
Although this calculation is performed using the radiation era background of the \AMR{} simulation, for illustrative purposes we present these results on \Fref{fig:isotropic} over a range of redshifts between recombination and today, and compare with a similar calculation performed using $1000$ \LCM{} realizations in the \LCDM{} cosmology.  
In order to map the $\beta_f$ derived from the \AMR{} pseudo-realizations onto \Fref{fig:isotropic}, we perform two re-scalings.  
First, we observe that the duration of the \AMR{} simulation from $\eta_i$ to $\eta_f$ corresponds to a growth factor of $\eta_f / \eta_i$.  
Using the same growth factor in the \LCDM{} cosmology defines a redshift $z_f$ such that 
\begin{align}
    \frac{\eta_{\LCDM{}}(z_f)}
         {\eta_{\LCDM{}}(z_\mathrm{rec})}
    =
    \frac{\eta_f}{\eta_i} 
    \;.
    \label{eq:lcm_eta_ratio_mapping}
\end{align}
For example, the radiation-era interval \(17\to117\) has \(\eta_f/\eta_i=117/17 \approx 6.88\).  
Taking \(z_\mathrm{rec}=1100\) and using the full \LCDM{} expansion history gives \(z_f \simeq 52.4\).  
Second, we observe that the \LCM{} predicts $\beta \propto \sqrt{\xi_d}$, so we rescale the \AMR{} value 
\begin{align}
    \beta_f &\to
    \sqrt{
    \frac{\xi_d^{\LCDM{}}(z_{\rm rec}\rightarrow z_f)}
    {\xi_d^\mathrm{rad}(\eta_i\rightarrow\eta_f)}
    }\,\beta_f
    \;,\quad{\rm where}\nonumber\\
    \xi_d^{\LCDM{}}(z_{\rm rec}\rightarrow z_f)
    &:=
    \frac{
    \displaystyle\int_{\eta(z_{\rm rec})}^{\eta(z_f)}
    d\eta\,
    \frac{[\eta(z_f)-\eta]^2}{\eta^3}\,
    \xi_d^{\LCDM{}}[z(\eta)]
    }{
    \displaystyle\int_{\eta(z_{\rm rec})}^{\eta(z_f)}
    d\eta\,
    \frac{[\eta(z_f)-\eta]^2}{\eta^3}
    }\;,
    \nonumber\\
    \xi_d^\mathrm{rad}(\eta_i\rightarrow\eta_f)
    &:=
    \frac{
    \displaystyle\int_{\eta_i}^{\eta_f}
    d\eta\,
    \frac{(\eta_f-\eta)^2}{\eta^3}\,
    \xi_d^\mathrm{rad}(\eta)
    }{
    \displaystyle\int_{\eta_i}^{\eta_f}
    d\eta\,
    \frac{(\eta_f-\eta)^2}{\eta^3}
    }\;,
\end{align}
using the extrapolated string density parameter from \Eref{eq:xid_lcdm_extrapolated} in the numerator, and using the radiation era parameter from \Eref{eq:xid_four_xit_rad} in the denominator. 

\Fref{fig:isotropic} shows the mean and standard deviation of $\beta$ for each of $100$ \AMR{} pseudo-realizations and $1000$ \LCM{} realizations.  
We find that the standard deviations are in reasonably good agreement between these two methods across a range of times, although $\beta_\mathrm{rms}$ is consistently smaller for \AMR{}.  
This level of agreement can be taken as an indication of the reliability of the calibration.  
The smaller values for \AMR{} may be a consequence of correlations between the pseudo-realizations.  
Unlike the \LCM{} realizations that are completely independent of one another, the \AMR{} pseudo-realizations are correlated, because they are constructed from overlapping regions of the same cubic simulation volume.  
These correlations should tend to reduce the spread in $\beta$, leading to a smaller standard deviation.  
Since we only use independent \LCM{} realizations for making predictions of isotropic birefringence today, these correlations do not impact our conclusions regarding the consistency of axion-string induced birefringence and reported measurements of isotropic birefringence. 

\section{Statistics and limit setting}
\label{sec:stats}

This section describes how we use measurements of the birefringence angular power spectrum to set limits on $\mathcal{A}$.  
\begin{enumerate}
    \item  We calculate $C_\ell^{\alpha\alpha}$.  After calibrating the \LCM{} using the \AMR{} simulations, there is only a single free parameter $\theta = \mathcal{A}^2$, which corresponds to an $\ell$-independent scaling.  The predicted angular power spectrum is written as $C_\ell^{\alpha\alpha} = \theta \, t_\ell$ where $t_\ell = C_\ell^{\alpha\alpha} |_{\mathcal{A}^2=1}$ is the template. 
    \item  The various \CMB{} telescopes report measurements of bandpowers, which correspond to a weighted average of $C_\ell^{\alpha\alpha}$ over a bin of multipoles.  Each bin has a range $\ell_\mathrm{min} \leq \ell \leq \ell_\mathrm{max}$ and a midpoint $b = (\ell_\mathrm{min} + \ell_\mathrm{max} + 1) / 2$.  Denote the measurements as $\bar{C}_{b} \pm \Delta C_{b}$.  The bandpowers are reported in 
    Table~3 of \Rref{Bortolami:2022whx} for Planck 2018, 
    Fig.~7 of \Rref{BICEPKeck:2022kci} for BICEP / Keck 2018, 
    Table~3 of \Rref{SPT:2020cxx} and \Rref{SPTpol:webpage} for SPTpol 500d,
    and Fig.~3 of \Rref{Namikawa:2020ffr} and \Rref{ACTpol:webpage} for ACTpol. 
    \item  We calculate model bandpower by taking the uniform-per-mode average of $t_\ell$ for each multipole in the bin: 
\begin{align}
    t_b = \frac{\sum_{\ell = \ell_\mathrm{min}}^{\ell_\mathrm{max}} ( 2 \ell + 1 ) \, t_\ell}{\sum_{\ell = \ell_\mathrm{min}}^{\ell_\mathrm{max}} ( 2 \ell + 1 )} 
    \;.
\end{align}
    Planck 2018 measurements are reported at every multipole up to $\ell = 24$, and no binning is needed.  
    SPTpol uses an optimal $w_L = C_L^{\rm theory}/\mathrm{Var}$ weighting and ACTpol/BK18 do not state their intra-bin weighting; for those three, our uniform-per-mode weighting is an approximation whose leading uncertainty is the lowest-bin weight.
    \item  We construct the Gaussian bandpower likelihood 
    \begin{align}
    \mathcal{L}(\theta) = \mathrm{exp}\bigl[ - \chi^2 / 2 \bigr]
    \qquad \text{where} \qquad 
    \chi^2 = \sum_{b} \frac{\bigl( \bar{C}_{b} - \theta t_b \bigr)^2}{(\Delta C_{b})^2} 
    = \frac{(\theta - \hat{\theta})^2}{\sigma_\theta^2} 
    \;.
    \end{align}
    Since the log-likelihood is quadratic, the maximum-likelihood estimate $\hat{\theta}$ and its Gaussian error $\sigma_\theta$ are calculated: 
    \begin{align}
    \frac{\hat{\theta}}{\sigma_\theta^2} = \sum_b \frac{t_b \, \bar{C}_b}{(\Delta C_{b})^2} 
    \qquad \text{and} \qquad 
    \frac{1}{\sigma_\theta^2} = \sum_b \frac{t_b^2}{(\Delta C_{b})^2} 
    \;.
    \end{align}
    \item  We set a flat prior $p(\theta) = \Theta(\theta)$, which enforces $\theta \geq 0$ and implies that the posterior is a Gaussian truncated at zero.  We define the $C$-level (\eg{}, $C = 0.95$) upper limit on $\theta = \mathcal{A}^2$ to be the solution of 
    \begin{align}
    C = \frac{\int_{-\infty}^{\theta_C} \! \dd \theta \ p(\theta) \, \mathcal{L}(\theta)}{\int_{-\infty}^{\infty} \! \dd \theta \ p(\theta) \, \mathcal{L}(\theta)} 
    \qquad \Leftrightarrow \qquad 
    \theta_C = \hat{\theta} + \sigma_\theta \, \Phi^{-1}\bigl[ 1 - (1-C) \, \Phi[\hat{\theta}/\sigma_\theta] \bigr]
    \;,
    \end{align}
    where $\Phi(x)$ is the standard-normal CDF and $\Phi^{-1}(p)$ is its inverse.  
    We calculate the upper limit on $|\mathcal{A}|$ as 
\begin{align}\label{eq:Acal95}
    |\mathcal{A}| < \mathcal{A}_{95} = \sqrt{\theta_{95}} \quad \text{at 95\% C.L.}
    \;.
\end{align}
    \item  To account for the extrapolation uncertainty, discussed in \Sref{sub:extrapolation_uncertainty}, we introduce a multiplicative nuisance parameter \(\kappa\), which scales the template uniformly across multipoles, \(t_\ell\to\kappa t_\ell\).  The \VOS{}-motivated tests in \Fref{fig:app_vos_xid_uncertainty} show that alternative late-time \(\xi_d\) histories mainly change the normalization of the \LCM{} spectrum, with only a mild tilt.  We therefore model the residual extrapolation uncertainty as an order-one normalization uncertainty.  We take a log-normal prior
\begin{align}
    p(\kappa)
    \propto
    \frac{1}{\kappa}
    \exp\left[
    -\frac{(\ln\kappa)^2}{2s^2}
    \right],
    \qquad
    s=\ln2,
\end{align}
    such that \(68\%\) of the prior mass lies in the range \(\kappa\in[1/2,2]\).  We define the \(C\)-level upper limit on \(\theta=\mathcal{A}^2\) to be the solution of
\begin{align}
    C =
    \frac{
    \int_0^\infty \dd\kappa\,p(\kappa)
    \int_{-\infty}^{\theta_C} \dd\theta\,p(\theta)\,
    {\cal L}(\kappa\theta)}
    {
    \int_0^\infty \dd\kappa\,p(\kappa)
    \int_{-\infty}^{\infty} \dd\theta\,p(\theta)\,
    {\cal L}(\kappa\theta)}
    \,,
\end{align}
where
\begin{align}
    {\cal L}(\kappa\theta)
    =
    \exp\left[
    -\frac{(\kappa\theta-\hat{\theta})^2}
          {2\sigma_\theta^2}
    \right]
    \;.
\end{align}
We calculate the 95\% C.L. upper limit on $\mathcal{A}$ as in \Eref{eq:Acal95}.  
All of the limits reported in the main article (\ie{}, values of $\mathcal{A}_{95}$) have taken into account the extrapolation uncertainty.  
\end{enumerate}
We perform this procedure for each of the \CMB{} telescopes discussed in the main text.  
Since covariances are not available, our statistical treatment neglects the covariance and assumes that the measurements in different bins are uncorrelated. 
Similarly without covariances we cannot implement the Hamimeche-Lewis likelihood \cite{Hamimeche:2008ai} used by the telescopes, and instead we adopt a Gaussian likelihood.  
For each telescope we find that the limit is driven by the lowest-multipole bin, which suggests that these results are robust against introducing inter-bin covariance.

For visual guidance only, \Fref{fig:anisotropic} shows one-sided \(95\%\) upper limits for each telescope's bandpowers.  
For each bin \(b\), we approximate the reported uncertainty by a Gaussian distribution centered at zero and define \(C_{b,95}\) through
\begin{align}\label{eq:illustrative_bandpower_limit}
    0.95
    & =
    \int_{-\infty}^{C_{b,95}}
    \frac{\dd C_b}{\sqrt{2\pi}\,\Delta C_b}
    \exp\!\left[-\frac{C_b^2}{2(\Delta C_b)^2}\right] 
    \qquad \Rightarrow \qquad 
    C_{b,95} \approx 1.645\,\Delta C_b
    \;.
\end{align}
The downward arrows in \Fref{fig:anisotropic} show these limits, converted to the plotted \(D_b=\ell_b(\ell_b+1)C_b/(2\pi)\) convention at the effective multipole of each bin. The horizontal bars indicate the multipole range of the corresponding bandpower bin. They are intended only as an illustration of the approximate bandpower sensitivity; the limits on \(\mathcal A\) are obtained from the full procedure described above. 

\end{document}